\documentclass[review]{elsarticle}
\usepackage{amsmath,amssymb}
\usepackage{subfigure}
\usepackage{caption}    
\usepackage{color}
\usepackage{url}
\usepackage{lineno,hyperref}
\usepackage{graphicx}
\usepackage{dcolumn}
\usepackage{bm}
\modulolinenumbers[5]
\pdfoutput=1
\journal{Computers \& Fluids}

\begin{document}

\begin{frontmatter}

\title{A unified gas kinetic scheme for transport and collision effects in plasma}

\author[mymainaddress,mysecondaryaddress]{Dongxin Pan}
\ead{chungou@mail.nwpu.edu.cn}

\author[mysecondaryaddress]{Chengwen Zhong}
\ead{zhongcw@nwpu.edu.cn}

\author[mysecondaryaddress]{Congshan Zhuo\corref{mycorrespondingauthor}}
\cortext[mycorrespondingauthor]{Corresponding author}
\ead{zhuocs@nwpu.edu.cn}

\author[mymainaddress]{Wei Tan}

\address[mymainaddress]{State Key Laboratory of Astronautic Dynamics, Xi'an, Shaanxi 710043, China. }
\address[mysecondaryaddress]{National Key Laboratory of Science and Technology on Aerodynamic Design and Research, Northwestern Polytechnical University, Xi'an, Shaanxi 710072, China.}

\begin{abstract}
In this study, the Vlasov-Poisson equation with or without collision term for plasma is solved by the unified gas kinetic scheme (UGKS). The Vlasov equation is a differential equation describing time evolution of the distribution function of plasma consisting of charged particles with long-range interaction. The distribution function is discretized in discrete particle velocity space. After the Vlasov equation is integrated in finite volumes of physical space, the numerical flux across a cell interface and source term for particle acceleration are computed to update the distribution function at next time step. The flux is decided by Riemann problem and variation of distribution function in discrete particle velocity space is evaluated with central difference method. A electron-ion collision model is introduced in the Vlasov equation. This finite volume method for the UGKS couples the free transport and long-range interaction between particles. The electric field induced by charged particles is controlled by the Poisson's equation. In this paper, the Poisson's equation is solved using the Green's function for two dimensional plasma system subjected to the symmetry or periodic boundary conditions. Two numerical tests of the linear Landau damping and the Gaussian beam are carried out to validate the proposed method. The linear electron plasma wave damping is simulated based on electron-ion collision operator. Compared with previous methods, it is shown that the current method is able to obtain accurate results of the Vlasov-Poisson equation with a time step much larger than the particle collision time. Highly non-equilibrium and rarefied plasma flows, such as electron flows driven by electromagnetic field, can be simulated easily.
\end{abstract}


\end{frontmatter}


\section{Introduction}
\label{Introduction}
The Vlasov equation describes time evolution of the distribution function of plasma which consists of charged particles with long-range interaction \cite{vlasov1938vibration}. Vlasov showed the difficulties in description of plasma with long-range Coulomb interaction when kinetic theory based on standard transport-collision is applied: 1) Theory of pair collisions disagrees with the discovery by Rayleigh, I. Langmuir and L. Tonks that vibrations exist in electron plasma; 2) Theory of pair collisions without Coulomb interaction will lead to divergence of kinetic term; 3) Theory of pair collisions cannot explain results of experiments by H. Merrill and H. Webb that electrons scatter anomalously in gaseous plasma \cite{merrill1939electron}. In gas, binary interaction is taken as the rule. However, in plasma, waves, or organized motion of plasma, are very important because the particles can interact at long ranges through the electric and magnetic forces. Vlasov suggested a source term which contains the Coulomb interaction added in collisionless Boltzmann equation. The Vlasov equation is a partial differential equation (PDE) for distribution function of particles, which represents probability that particle stays at a specific position and velocity. The acceleration term in equation describes redistribution of particle on particle phase space due to electromagnet force.

The microscopic method is a way to simulate particle motion directly. The trajectory of individual particle is traced and the distribution of particle is quantified with the number of particles per unit volume and per velocity interval. The macroscopic variables can be obtained by moment results of microscopic variables. P. Degond et al(2010) proposed a particle-in-cell (PIC) method based on Vlasov equation \cite{degond2010asymptotic}. Individual particles in a Lagrangian frame are traced in physical space and particle velocity space. Under the Eulerian frame of reference, the mass and momentum conservation equations are obtained with velocity moments of the Vlasov equation \cite{degond2010asymptotic}. In this method, the particle details, such as the location, velocity, are exactly captured by the equation of motion. The PIC method also has a advantage on multi-component problems in plasma simulations. The biggest issue of PIC is the huge computational cost when the number of particles increases. For this reason, its application is limited in problems which consist of a small number of particles.

The finite element method (FEM) is used for finding approximate solutions for distribution function in the Vlasov equation. In FEM the whole computational domain divided into several discrete elements. A Polynomial is applied to approximate the distribution function in phase space. For every element a group of equations is built. These finite elements are then assembled into a larger system of equations which can be solved to obtained the distribution functions. N. Crouseilles et al(2011) proposed a Galerkin method for the Vlasov equation. Lagrange polynomials are applied to construct basis function. A system of linear equations is constructed after the Vlasov equation is discretized in all cells. Solution of this linear equation is approximation for unknown distribution functions of the particles \cite{crouseilles2011discontinuous}. D. C. Seal(2012) reported a work in his doctorate thesis about a discontinuous Galerkin method for solution of the Vlasov equation. In finite element space, the basis functions used for representation of the distribution function in the Vlasov equation are allowed to be discontinuous at the cell interface. The Legendre polynomials which contain unknown coefficients are applied to discretize the Vlasov equation. The discrete equation is integrated over mesh cells. The unknown coefficients in basis functions are obtained solving the advection equation. At cell interface, the evaluation of numerical flux is turn into a Riemann problem \cite{seal2012discontinous}. R. Heath et al.(2012) proposed a upwind penalty Galerkin method for approximating the Vlasov-Poisson system. Piecewise constant functions are used as basis functions. The use of penalty function preserves the positivity of the electron distribution functions. The convergence is also greatly improved \cite{heath2012discontinuous}. To determine the coefficients in polynomial about phase space, the particle distribution in physical space and velocity space is needed and equation of particle motion is solved at each time step. A rather high computational cost still becomes a barrier for plasma simulation.

The spectral method is a class of methods which can be used to solve the Vlasov equation. The idea is to construct a series of basis functions. The sum of these functions is the solution of the Vlasov equation. Over the whole domain, the basis functions keep nonzero, which is the main difference from FEM. J. W. Schumer et al.(1998) constructed a spectral method using the Fourier-Hermite polynomial to discretize this equation in phase space. They finally obtained a system of equations which can be solved at every time step using Runge-Kutta method (RK). However, this method does not show a good convergence behavior. It is not a robust scheme. Only fourth Order RK can finally reach convergence \cite{schumer1998vlasov}. S. Le Bourdiec et al(2006) constructed a spectral method to solve the Vlasov-Poisson system using generalized Hermite functions \cite{le2006numerical}. The distribution function is expanded at the Nth-spectral order of approximation with Hermite polynomial. So, the problem is finally recast in a finite system of hyperbolic equations. This method can produce rather accurate results if the Hermite polynomial is properly applied. But truncation error from Hermite polynomial approximation has a great effect on numerical results \cite{le2006numerical}. N. Crouseilles et al.(2009) proposed a forward semi-Lagrangian spectral method. The distribution function is projected on B-splines basis. The particle trajectory is used to decide the coefficient in the sum of basis functions. This method has a good robustness and allows a very large time step. However, because of lack of an effective reconstruction, the conservation is slightly lost, which causes a bit loss of accuracy \cite{crouseilles2009forward}. The spectral methods have lower computational cost than FEM. But in problems with complex geometries, it becomes less accurate.

Conservative methods for solution of the Vlasov equation are also paid great attention. E. Sonnendr¨¹cker(1999) et al. first introduced a semi-Lagrangian method. Different from the PIC methods which traces a finite number of particles following characteristic curves and describe interaction of particles with the electric fields on a grid, this explicit conservative method computes distribution function on grid by following the characteristic curves. The reconstruction at a cell interface is realized along characteristic lines which contain grid point values of the distribution function. In this method, the spline interpolation is used to reconstruct the distribution function at the cell interface. The coefficients in polynomial are decided by the characteristic curves, which are obtained by equation of particle motion \cite{sonnendrucker1999semi}. F. Filbet et al (2001) constructed a particle based conservative method. As in the finite volume method, the particle location and the velocity are traced and the distribution function in each cell is computed. A slope corrector is used in reconstruction of distribution function at the cell interface to ensure the preservation of positivity and the maximum principle. The flux is evaluated by solving the Riemann problem. This method is very robust and time step is not restricted by a CFL condition \cite{filbet2001conservative}. The accuracy is severely influenced by the phase and amplitude error because of dissipation from the way of reconstruction and the use of slope corrector. N. Crouseilles et al (2007) constructed a conservative method using the Hermite spline interpolation, which is used for expansion of distribution function \cite{crouseilles2007hermite}.  The coefficient of polynomial is decided by the particle characteristics. The equation of particle motion in electric field is discretized in time and solved in phase space. The distribution function can be obtained by the summation of the series after solving a linear system \cite{crouseilles2007hermite}. The time step is restricted by the scale of particle motion and the computational cost does not show an advantage over PIC methods. These conservative methods partly improve the efficiency of PIC method and the conservative property. But they still fail to describe real particle motion in phase space with a low computational cost.

In recent years, the kinetic theory of gases is being studied to analyze the particle dynamics in plasma. At the microscopic level, particles are moving and interacting with each other randomly, but a collection of particles can be described with deterministic macroscopic properties. Kinetic theory analyzes the particle dynamics based on statistics, which links between the microscopic and macroscopic evolutions together. This method avoids huge computational cost since it does not need to trace individual particle. For a system which consists of a certain number of particles, a point in phase space represents a dynamic state of each particle. J. W. Banks et al. (2011) proposed a finite volume method (FVM) to solve the Vlasov equation. the PDE is integrated in discrete phase space. The numerical flux of the distribution function through a cell interface is evaluated using high-order approximations of the cell-face average \cite{banks2011two}. S. Xu et al (2015) simulated the NS-DBD plasma formation between parallel electrodes in $N_2-O_2$ mixture air at low-pressure under nanosecond impulses \cite{Xu2016Modeling}. Based on the kinetic model, the complex physical process of nanosecond-pulse electrical discharges including the variations of multiple physical parameters are accurately captured. The defect of this method is the way of reconstruction at the cell interface. The characteristic curves for particle transport are lost in flux evaluation. An artificial approximation has replaced the physical situation at the cell interface. So a wrong dissipation is brought in because of the inaccurate reconstruction for particle state at the cell interface.

In current paper, a unified gas kinetic scheme (UGKS) is applied to simulate plasma flows based on the Vlasov equation. The UGKS is proposed by Xu, Huang and Yu \cite{xu2010unified,huang2012unified}, which attracts more and more researchers' attention \cite{xu2011improved,chen2012unified,liu2012modified,wang2012study,mieussens2013asymptotic,liu2014unified,liu2014investigation,liu2015unified,chen2015comparative,sun2015asymptotic,liu2016unified,liu2016asymptotic,zhu2016implicit,zhu2017unified}. In continuous phase space, the Maxwellian distribution function can be easily used to integrate the Vlasov equation in a finite volume. But in plasma flow, we cannot make an assumption that particles stay in an equilibrium state. The real gas distribution function in the highly non-equilibrium region is never able to be described by a Maxwellian velocity distribution. In this paper, the particle velocity space is discretized so that distribution can be exactly described in all flow regimes \cite{xu2010unified}. The flux reconstruction at the cell interface is treated as the Riemann problem along characteristic curves for particle transport \cite{xu1998karman}. The free transport and interaction between particles are coupled so that the dissipation in the transport process is controlled by the source of the long-range interaction rather than numerical time step. With this technique, the artificial dissipation as the one in Ref.~\cite{banks2011two} will not be brought in numerical process.

For computational cost, the UGKS is a much more efficient method than PIC method. The PIC method includes the four typical procedures: 1) Equation of particle motion is integrated; 2) Properties of particles are interpolated to field mesh; 3) Field is reconstructed on mesh points; 4) Flow field is interpolated from mesh to particle locations. The information is communicated between particles and macroscopic field at every time step. In real system being studied, the number of particles is often extremely large, resulting in poor communication efficiency. The simulation becomes inefficient and even impossible at all. Different from the way of description for flows in PIC method, the UGKS describes flow motion using particle distribution function. It gives the number density of particles in the six-dimensional phase space. This method is a statistical method based on particle motion. Instead of following individual particle, the distribution function is used to show probability of particles which locate in a spatial point and a certain velocity interval. The macroscopic variables can be obtained by the integral of the microscopic variables in particle velocity space. Internal motions of particle, such as rotation and vibration, are taken into account through extra internal degrees of freedom \cite{xu1998karman}. The UGKS avoids huge computational cost produced by solving the equation of particle motion and tracking particle as stated in PIC method. This high fidelity and efficiency in numerical simulation make the UGKS a better way to solve plasma problems. Based on the BGK model equation, C. liu et al(2017) simulated multi-scale and multi-component plasma transport based on the UGKS \cite{Liu2017A}. Many challenging cases including magnetic reconnection problem in the transition regime are simulated. The UGKS is a physically reliable multi-scale plasma simulation method which provides a powerful and unified approach for the study of plasma physics.

In solving the Vlasov equation, the UGKS can overcome the negative effect of truncation error caused by the approximation of the distribution function in FEM. In FEM the phase space is divided into a collection of sub-domains. A set of equations are used to represent discretized PDE in every subdomain. All sets of element equations are finally combined for solution of the whole discrete phase space. In construction of subdomains, the distribution function is approximated by polynomial. FEM tries to minimize the truncation error from this approximation. So, integration of inner product of residual and weight functions is set to zero. In this procedure, error is caused by trial functions for distribution function in every subcell of phase space. In the UGKS, the Vlasov equation is integrated and the flux is evaluated at cell interface of phase space. This flux is obtained by the solution for the local Riemann problem \cite{godunov1959difference}. At the cell interface, the direction of propagation of particle information is simulated. And then the flux is reconstructed by using difference biased in the direction determined by the sign of the characteristic speeds \cite{courant1952solution}. The particle velocities propagated to cell interface are the characteristics in the UGKS, which gives a correct representation for physical motion. The only source of error comes from the dissipation which depends on the resolution of a mesh cell.

The UGKS also has a more concrete physical foundation than spectral method. Spectral method writes distribution function in the Vlasov equation as a sum of basis functions. The solution must satisfy the control equation as well as possible so coefficients before basis should be carefully chosen. In the case that the solution is smooth, the spectral method has a high order accuracy \cite{canuto2007spectral}. However, a highly non-equilibrium state will challenge this method because a large error appears in a strong discontinuity at cell interface of phase space. The approximation using polynomial can not achieve a fidelity when the distribution function deviates from the Maxwellian distribution in non-equilibrium state. In physics, the distribution function can have arbitrary geometric features in plasma so polynomial often fails to capture actual distribution in phase space. In equilibrium state, the Vlasov equation can be integrated under continuous particle velocity space in the flux evaluation based on FVM. Because in this regime, analytical formulas for gas distribution function is well-defined. However, a plasma has properties unlike those of fluid flow. Although particles are unbound, their transport is not entirely free. When a charged particle is moving, time-varying electric field is produced around it. In plasma, the movement of a charged particle affects and is affected by the general field created by the movement of other charged particles. This process  causes collective behavior with many degrees of variation \cite{sturrock1994plasma,hazeltine2004framework}. Collision interactions are often weak in hot plasmas and external forcing can drive the plasma far from local equilibrium and lead to a significant population of unusually fast particles. For these reason, we can not have a confidence that we can still approximate distribution function of particles in plasma under continuous velocity space. In the UGKS, the particle velocity space is discretized into a number of points which represent every dynamical state. The Vlasov equation is split into PDEs for distribution function at each velocity point. In fact, we solve the Vlasov equation for distribution function at each point of discrete phase space. So, a high fidelity is obtained when distribution function is solved at each time step, which is the advantage over spectral method. Theoretically, the physical content in the UGKS is far too richer than methods based on numerical approximation using polynomials.

Unlike neutral gas, plasma consists of a significant number of charge carriers, which makes it electrically conductive. Because of this feature, plasma responds strongly to electromagnet field. Plasma does not have certain shape or certain volume if not enclosed in a container, but its motion and state are able to be controlled by electromagnet force. Under influence of electromagnet field, plasma can be formed into various structures. An important property used to describe the electric field is electric potential. It denotes the amount of electric energy a unitary point charge have when located at any point in electric field. The electric potential can be obtained by work done by an external agent in carrying a charge from reference point to present location. The force and electric potential are directly related. The electric potential of a point charge declines as it is driven by an electric force and moving in the direction of electric field line. By Faraday's law, the electric field has zero curl. So electric field can be obtained directly by electric potential. The Maxwell equations are a set of PDEs that describe how the fields vary in space due to sources. For simplification of computation, under two dimensional configuration, only electric field is consideration in simulation of plasma flows. In this work, we are going to solve equation for description of electric field instead of complete Maxwell equation. In this group of equations, the electric potential is described by the Poisson's equation. In mathematics, electric field leaving a volume is proportional to the charge inside. The Green's function is applied to solve the Poisson's equation which denotes electric potential. Electric potential obtained by the Poisson's equation is a sum of series whose basis functions are Green's functions. The basis functions represent Coulomb interaction potential induced by charge in the whole computational domain \cite{mouhot2010landau}. After the electric potential is obtained, the electric field can be obtained by spatial gradient of the electric potential. The acceleration term in the Vlasov equation can be decided.

The rest of the paper is organized as follows. First, we emphasize the numerical method for solution of the Vlasov-Poisson equation in Section \ref{sec:Numerical methods}. Then, the Landau damping and the Gaussian beam are numerically simulated and the numerical results are presented in Section \ref{sec:Numerical results}. Finally, some remarks concluded from this study are grouped in Section \ref{sec:Conclusions}.

\section{Numerical methods}\label{sec:Numerical methods}
The plasma Vlasov equation can be written as
\begin{equation}\label{Eq01}
\frac{{\partial f}}{{\partial t}} + {\bm{u}} \frac{{\partial f}}{{\partial {\bm{x}}}} + {\bm{a}} \frac{{\partial f}}{{\partial {\bm{u}}}} = 0,
\end{equation}
where $f$ is the particle distribution function of the space $\bm{x}$, time $t$, particle velocity $\bm{u}$, and acceleration $\bm{a}$ due to electromagnetic field. The relation between the macroscopic variables $\bm{W}$ with the microscopic variables $\bm{\psi}$ is
\begin{equation}\label{Eq02}
{\bm W} = \left( \begin{array}{c}
\rho \\
\rho U\\
\rho V\\
\rho E
\end{array} \right) = \int {{\bm \psi} fdudvd\xi },
\end{equation}
where $\rho E = \frac{1}{2}\rho \left( {{U^2} + {V^2} + \frac{{K + 2}}{{2\lambda }}} \right)$, and ${\bm \psi} =(1, u, v, 1/2(u^2+v^2+\xi ^2))^T$. The pressure and density can be related with $\lambda $ as  $p = {\rho  \mathord{\left/ {\vphantom {\rho  {2\lambda }}} \right. \kern-\nulldelimiterspace} {2\lambda }}$. $\bm u = (u,v)$ is the particle velocity and $\xi$ is the internal freedom of gas.

In discrete particle velocity space, Eq.~(\ref{Eq02}) can be written as
\begin{equation}\label{Eq03}
\bm{W} = \sum { \sum {\sum {\bm{\psi} f{\Delta u}{\Delta v}\Delta \xi } } }.
\end{equation}

The particle velocity $\bm u $ is discretized with intervals ${\Delta u}$ and ${\Delta v}$. In finite volume, we integrate the Vlasov equation in physical space for each discrete velocity \cite{liu2012modified}. The result of integration can be written as
\begin{equation}\label{Eq04}
f_{i,j,k}^{n + 1} = f_{i,j,k}^n + \frac{1}{\Omega _{i}}\int_0^{\Delta t} {\sum\limits_p^q \bm{u}  \cdot \bm{S}_p
{f_{cf,i,j,k}}dt }  -{ \left( \bm{a} \cdot \frac{{\partial {f^n}}}{{\partial \bm{u}}} \right)_{i,j,k}}\Delta t,
\end{equation}
where $i$ is the cell index of a unstructured mesh, $j$ and $k$ present the index of particle velocity. The distribution function at the $n + 1$ time step can be obtained through Eq.~(\ref{Eq04}).

In current work, the charged particles interact with each other by long-range electromagnet force. The advection term can be evaluated as numerical flux along characteristic curves. In the Vlasov model, The distribution function around the cell interface can be reconstructed as \cite{xu1999gas}
\begin{equation}\label{Eq05}
{f_{cf,i,j,k}} = \left\{ \begin{array}{l}
f_{i,j,k}^l \qquad \bm{u} \cdot {{\bm{S}}_p} \ge 0,\\
f_{i,j,k}^r \qquad \bm{u} \cdot {{\bm{S}}_p} < 0.
\end{array} \right.
\end{equation}

The electromagnet force term is obtained by central difference method. For a dimension, difference between $f$ and particle velocity $u$ and $v$ can be written as
\begin{equation}\label{Eq06}
\begin{aligned}
\left ( \frac{{\partial f}}{{\partial u}} \right)_i &= \frac{{{f_{i, j + 1,k}} - {f_{i, j - 1,k}}}}{{2\Delta u}}, \\
\left ( \frac{{\partial f}}{{\partial v}} \right)_i  &= \frac{{{f_{i, j,k + 1}} - {f_{i, j,k - 1}}}}{{2\Delta v}}. \\
\end{aligned}
\end{equation}

At the boundary of particle velocity space for two dimensional plasma, we have
\begin{equation}\label{Eq07}
\begin{aligned}
{f_{i, - 1,k}} &= 0, \\
{f_{i, Nu + 1,k}} &= 0, \\
{f_{i, j, - 1}} &= 0, \\
{f_{i, j,Nv + 1}} &= 0, \\
\end{aligned}
\end{equation}
where $Nu$ and $Nv$ are the total numbers of discrete particle velocities in each dimension.

Next, we will introduce the process of solving the electromagnet Poisson's equation. Solving the Poisson's equation amounts to finding the electric potential $\Phi$ for a given charge distribution. The mathematical details behind the Poisson's equation in electrostatics are described by Gauss's law for electricity \cite{Maxwell1982A}. The Poisson's equation applied in our work is written as
\begin{equation}\label{Eq08}
\Delta \Phi  =  - \frac{{{\rho _c}}}{\varepsilon },
\end{equation}
where $\rho _c$  is the charge density and $\varepsilon$ is the permittivity of the medium. Eq.~(\ref{Eq08}) is solved using the Green's function \cite{christlieb2004efficient}. First, we find the particular solution.
\begin{equation}\label{Eq09}
{\Phi _i} = \frac{1}{\varepsilon }\sum\limits_{j = 1}^{Nc} {G\left( {{{\bm{x} }_i},{{\bm{x} }_j}} \right){Q_j}},
\end{equation}
where $Nc$ represents the number of cells, $i$ represents the cell being studied, $j$ represents cells in discrete computational domain and $j \ne i$. $Q_j$ represents charge in cell $j$. $G\left( {{{\bm{x} }_i},{{\bm{x} }_j}} \right)$ is the Green's function for the Poisson's equation, which can be written as
\begin{equation}\label{Eq10}
G\left( {{{\bm{x} }_2},{{\bm{x} }_2}} \right) = \frac{1}{4 \pi | {\bm{x}_1 - \bm{x}_2} | }.
\end{equation}

In the Green's function, $\bm{x}_1$ and $\bm{x}_2$ represent radius vectors for two locations of charges. To give a unique Green's function, symmetry or periodic boundary conditions should be implemented \cite{Sel2006Mathematical}. According to the Faraday's law, the electric field from charged particles is a conservative vector field. The electric potential $\Phi $ can be defined \cite{Griffiths2006Introduction}, such that
\begin{equation}\label{Eq11}
{\bm{E} _{self}} =  - \nabla \Phi .
\end{equation}

Eq.~(\ref{Eq11}) is used for electrostatic part of electric field. The electrodynamic part of electric field is induced by magnet field due to motion of charged particles. If magnet field is taken into consideration, the electric field induced by charged particles can be written as
\begin{equation}\label{Eq12}
{\bm{E} _{self}} =  - \nabla \Phi  - \frac{\partial \bm{A}}{\partial t},
\end{equation}
where $\bm{A}$ is the magnetic vector potential \cite{huray2011maxwell}. The relation between magnet field $\bm{B}$ and $\bm{A}$ can be written as
\begin{equation}\label{Eq13}
{\bm{B}} = \nabla  \times \bm{A}.
\end{equation}

The component of particle acceleration due to the electrostatic part of electric field obtained by particular solution via the Green's function and electrodynamic part of electric field induced by magnet field is computed as,
\begin{equation}\label{Eq14}
{\bm{a}} = - \frac{{{q_i}}}{m} \left( {\nabla \Phi  + \frac{{\partial \bm{A}}}{{\partial t}}} \right) = - \frac{{{q_i}}}{m}\left( {\frac{1}{\varepsilon }\sum\limits_{j = 1}^{Nc} {\nabla G\left( {{\bm{x}}_i},{{\bm{x} }_j} \right){Q_j}}  + \frac{\partial \bm{A}}{{\partial t}}} \right),
\end{equation}
where $q_i$ is the charge of the particle in cell $i$, $m$ is the mass of the particle.

Besides the electric field induced by the charged particles, the external electric field $\bm{E}_{appl}$ also contributes to variation of distribution function. So the acceleration term in Eq.~(\ref{Eq01}) includes two parts.
\begin{equation}\label{Eq15}
\bm{a} = \bm{a}_{self} + \bm{a}_{appl},
\end{equation}
where $\frac{q}{m}$ is charge to mass ratio. In this paper, the non-dimensional parameter satisfies relation $\left| {\frac{q}{m}} \right| = 1$. After substituting Eq.~(\ref{Eq05}), Eq.~(\ref{Eq06}) and Eq.~(\ref{Eq15}) into Eq.~(\ref{Eq04}), we can update the distribution function at $n + 1$ step. The above procedure can be repeated in the next time level.

\section{Numerical results}\label{sec:Numerical results}
\subsection{Linear Landau damping}
The Landau damping is the effect of damping of longitudinal space charge waves in plasma \cite{francis1984introduction}. It prevents an instability from developing and creates a region of stability in the parameter space \cite{lynden1962stability}. Energy exchange between electromagnetic wave in phase space and charged particles in plasma is the cause of Landau damping. In evolution, particles interact strongly with the wave \cite{tsurutani1997some}. In current work, we simulate the Landau damping based on the UGKS to study the way in which electrons interact and exchange energy with electromagnetic field. This case is used to test the accuracy of the UGKS for solving the Vlasov equation in describing time-variant electromagnetism parameters such as electric field and electric potential energy. First, an electromagnetic wave should be added to flow field. This wave is caused by an initial disturbance of distribution function \cite{landau1946vibrations}, which can be written as,
\begin{equation}\label{Eq16}
f = g{e^{ikx}},
\end{equation}
where $g$ is a Maxwellian distribution,
\begin{equation}\label{Eq17}
g = \rho \frac{\lambda }{\pi }{e^{ - \lambda \left( {{{\left( {u - U} \right)}^2} + {{\left( {v - V} \right)}^2}} \right)}}.
\end{equation}

In current work, 2D initial condition is set to
\begin{equation}\label{Eq18}
{f_0} = \rho \frac{\lambda }{\pi }{e^{ - \lambda \left( {{{\left( {u - U} \right)}^2} + {{\left( {v - V} \right)}^2}} \right)}}\left( {1 + \alpha \cos \left( {{k_x}x} \right)\cos \left( {{k_y}y} \right)} \right),
\end{equation}
where $\lambda  = 0.5$, $U = V = 0$ , $\alpha  = 0.05$ and the wave numbers ${k_x} = {k_y} = 0.5$ in current problem.

In the linear Landau damping case, a periodic boundary condition is applied. The four dimensional phase space contains $64$ points per dimension. The particle velocity is truncated at $6.0$. The boundary condition for 2D linear Landau damping is presented in Fig.~\ref{fig:Fig01}. The length for computational domain in each dimension is set to be $4 \pi$. The initial electronic density distribution is given in Fig.~\ref{fig:Fig02}. The constant time step $\Delta t$ is chosen as $0.01$. The evolution of density contour is presented in Fig.~\ref{fig:Fig03}. The energy of electric field is being carried away by electrons moving at the phase space. To show this process, the evolution of symmetric electric field and electric energy will be given. First, we computed the average of symmetric electric field. The logarithm scale of it is used for damping process of electric field, which is presented in Fig.~\ref{fig:Fig04}.

Now, the electric potential energy is computed to show the process of energy transport and conversion during interaction between the electric field and electrons. The electric potential energy is a potential energy that results from coulomb forces between charges. This kind of energy is associated with the configuration of a defined system which contains a certain number of charged particles. In physics, the electric energy of a system is the energy required for assembling charges from an infinite distance by bringing them close together. If an object has electric potential energy ${U_E}$, two key elements are crucial: 1) this object keeps its own charge; 2) it stays at a relative position to other electrically charged objects.
\begin{equation}\label{Eq19}
{U_E} = q\Phi .
\end{equation}

The evolution of potential energy in systems with time-variant electric fields is given in Fig.~\ref{fig:Fig05}.

In current work, we obtain the same evolution process of the electric field and electric energy as Ref.~\cite{filbet2001conservative}. Many works have been done to study the physical picture of Landau damping. Linearized theory is the simplest and rather complete way \cite{degond1986spectral}, but it is not physically reasonable because interaction between charged particles and electromagnet field is coupled with particle transport in the Vlasov equation. Linearized method is not appropriate to be used in this nonlinear system. However, nonlinearity has been a longstanding problem. To model nonlinear level in the Vlasov equation, a class of exponentially damped solutions of the Vlasov-Poisson equation is proposed \cite{caglioti1998time}. This method fails to explain the mechanism of energy transport although it approximates the damping curves in a mathematical way. The UGKS shows a great advantage on correctly representation for the Landau damping with a much lower computational cost than particle based methods.

\subsection{Nonlinear Landau damping}
In linear Landau damping case discussed above, the non-equilibrium phenomenon is still not very significant because the initial perturbation of density is very small. In this example, we increase amplitude of the initial perturbation of density in 1D space. The perturbation rate is set to be $\alpha  = 0.5$. The wave number and periodic length remain the same $k = 0.5$  and  $L = 4\pi $ respectively. The initial particle distribution function reads:
\begin{equation}\label{Eq220}
f = \frac{1}{{\sqrt {2\pi } }}{e^{ - \frac{{{u^2}}}{2}}}\left( {1 + \alpha \cos \left( {kx} \right)} \right).
\end{equation}

In UGKS, the discrete particle velocity space is applied. The 1D velocity space is truncated at $\left[ { - 6,6} \right]$ with $128$ intervals. When the velocity space is being discretized, we used cosine to decide the positions of discrete velocity points so that a high resolution can be obtained near the peak of distribution function. This equation is written as
\begin{equation}\label{Eq221}
{u_x} =  - {v_{\max }}\left( {1 + \cos \left( { - \frac{\pi }{2} + \pi \frac{i}{{Nu - 1}}} \right)} \right),
\end{equation}
where ${v_{\max }}$ is the maximum of particle velocity.

The evolution of the kinetic entropy is taken as a benchmark solution. Traditional FEMs always overestimate this value.The particle based methods may be an alternative, but computation is overwhelming. The UGKS is able to obtain the same accuracy as particle based method but calls for much lower computational cost. The comparison result between UGKS and Filbet et al \cite{filbet2001conservative} is presented in Fig.~\ref{fig:Fig06}. The Kinetic entropy is computed by $H =  - \sum {f\ln f} $.

Another benchmark result is ${L^2} - norm$ of the distribution function. Variation of ${L^2} - norm$ is often used to show the rate of negative values because the global mass $\int {fd\Xi } $ should be preserved. This value $\sum {{f^2}} $ is used to test the conservative characteristic of UGKS in simulation for plasma flow. The comparison result of ${L^2} - norm$ between UGKS and Filbet et al \cite{filbet2001conservative} is given in Fig.~\ref{fig:Fig07}.

The results of UGKS fit well with the ones of the particle based method. With a much lower computational cost, the UGKS has the same advantage in accuracy as particle based method. The UGKS has a good conservation property and good accuracy in reconstruction so that spurious dissipation is avoided. The nonlinear Landau damping case is a highly non-equilibrium phenomenon. To show this, particle distribution functions at different time steps at the center of space are presented in Fig.~\ref{fig:Fig08}.

Fig.~\ref{fig:Fig08} shows the distribution function in velocity space. To give a better description of particle distribution in phase space, the distribution function in the $\left( {x,u} \right)$  space is presented in Fig.~\ref{fig:Fig09}.

In nonlinear Landau damping case, the nonlinear effects are too important so previous traditional Galerkin methods always fail to simulate this highly non-equilibrium case accurately. In UGKS, the particle behavior is accurately obtained by solving the Vlasov equation with discrete particle velocity space.

\subsection{Gaussian beam}

In optics, the Gaussian beam is a monochromatic electromagnetic radiation which has a Gaussian intensity profile. As a result, it has a transverse magnet electromagnet field given by the Gaussian function. Such a beam can be expanded and focused through lens thus becomes different Gaussian beam at different time step. According to the theory of quantum mechanics, electrons can both interact with other particles and be diffracted. It has properties of particle and wave \cite{BroglieThe}. For Gaussian beam case, a beam of electrons is a good model since both interaction between particles due to electromagnet force and intensity of light should be simulated. In current paper, we choose a electron beam as model to be deformed by the electromagnet field. And the external electric field plays a role of lens. In evolution of beam, the external electric field has an opposite direction to electric field induced by electron itself. In the Vlasov equation, the acceleration term is from the electric field which includes two parts: Electric field ${\bm{E}} _{self}$ induced by electron given by the Poisson's equation and external electric field $\bm{E}_{appl}$. The external electric field is decided by self-consistent field method (SCF) using initial electric field induced by electrons. In current work, the external electric field is linear with respect to spatial coordinates. The relation between ${\bm{E}} _{self}$ and ${\bm{E}} _{appl}$ can be written as
\begin{equation}\label{Eq20}
\bm{E}_{self} + \bm{E}_{appl} =  - {\omega ^2}\bm{r},
\end{equation}
where $\omega$ represents the difference between the initial self-consistent electric field due to particle charges and the linear external electric field. Its value is decided by a self-consistent domain.
\begin{equation}\label{Eq21}
I = \left\{ {\left( {\bar x,\bar y,\bar u,\bar v} \right)|\frac{{{{\bar x}^2}}}{{{a^2}}} + \frac{{{{\bar y}^2}}}{{{a^2}}} + \frac{{{{\bar u}^2}}}{{{{\left( {\omega a} \right)}^2}}} + \frac{{{{\bar v}^2}}}{{{{\left( {\omega a} \right)}^2}}} = 1} \right\}.
\end{equation}

All values in Eq.~(\ref{Eq21}) are obtained by the root mean square (RMS) thermal velocity ${v_{th}}$. The RMS results for space and particle velocity are computed by
\begin{equation}\label{Eq22}
\begin{aligned}
\sqrt {{{\bar x}^2}}  &= \sqrt {\frac{{\int {{x^2}fd\Xi d\Omega } }}{{\int {fd\Xi d\Omega } }}}  = a,\sqrt {{{\bar y}^2}}  = \sqrt {\frac{{\int {{y^2}fd\Xi d\Omega } }}{{\int {fd\Xi d\Omega } }}}  = a, \\
\sqrt {{{\bar u}^2}}  &= \sqrt {\frac{{\int {{u^2}fd\Xi d\Omega } }}{{\int {fd\Xi d\Omega } }}}  = \omega a,\sqrt {{{\bar v}^2}}  = \sqrt {\frac{{\int {{v^2}fd\Xi d\Omega } }}{{\int {fd\Xi d\Omega } }}}  = \omega a.
\end{aligned}
\end{equation}

For Gaussian beam case, the initial self-consistent electric field is linear in space. So we can obtain the two unknown variables $\omega$ and $a$ using RMS thermal velocity $v_{th}$,
\begin{equation}\label{Eq23}
a = \frac{{{r_a}}}{2},{v_{th}} = \frac{{\omega {r_a}}}{2},
\end{equation}
where $r_a$ is the radius of beam. Now we introduce a tune depression factor $\chi  = \frac{{{\omega _0}}}{\omega }$ such that we can adjust external electric field with initial self-consistent electric field. The linear external electric field whose direction is opposite to initial self-consistent electric field can be written as
\begin{equation}\label{Eq24}
\bm{E}_{appl} = \omega _0^2\bm{r}.
\end{equation}

In evolution of a beam, the external electric field $\bm{E}_{appl}$ is used to focalize the beam. The dimensionless form of Eq.~(\ref{Eq01}) can be rewritten as
\begin{equation}\label{Eq25}
\frac{{\partial f}}{{\partial t}} + \bm{u}\frac{{\partial f}}{{\partial \bm{x}}} + \left( -  \left( {\bm{E}}_{self} + {\bm{E}}_{app} \right) \right) \frac{{\partial f}}{{\partial \bm{u}}} = 0.
\end{equation}

The negative sign $``-"$ is because of the negative charge carried by electron.

The 2D initial condition is written as
\begin{equation}\label{Eq26}
{f_0}{\rm{ = }}\left\{ {\begin{array}{*{20}{c}}
{\rho \frac{\lambda }{\pi }{e^{ - \lambda \left( {{{\left( {u - U} \right)}^2} + {{\left( {v - V} \right)}^2}} \right)}}}&{{x^2} + {y^2} \le r_a^2},\\
0&{{x^2} + {y^2} > r_a^2}.
\end{array}} \right.
\end{equation}

In kinetic theory, $\lambda$ can be obtained by \cite{xu2001a}
\begin{equation}\label{Eq27}
\lambda  = \frac{m}{{2{k_B}T}},
\end{equation}
where $m$ is the particle mass, ${k_B}$ is the Boltzmann constant, $T$ is the temperature. The mesh for beam case has a round boundary, which is shown in Fig.~\ref{fig:Fig10}. $4032$ mesh cells are used to discretize the computational domain.

Absorbing boundary condition (ABC) is applied in Gaussian beam case. The buffer zone is added between the computational domain and boundary with a range ${0.5^2} < {x^2} + {y^2} < {1.0^2}$. The flux damps along warp of a circular zone in buffer zone.
\begin{equation}\label{Eq28}
{F_{{r_o}}} = \delta {F_{{r_i}}},
\end{equation}
where $\delta$ is damping rate, ${r_o}$ and ${r_i}$ are the two location on a warp line respectively. The damping rate can be computed as
\begin{equation}\label{Eq29}
\delta  = \cos \left( {\frac{\pi }{2}\frac{{r - {r_i}}}{{{r_o} - {r_i}}}} \right).
\end{equation}

${r_i}$ and ${r_o}$ represent the two limit of buffer zone. In our work, we have
\begin{equation}\label{Eq30}
\begin{aligned}
{r_i} &= 0.5, \\
{r_o} &= 1.0.
\end{aligned}
\end{equation}

The location for ${F_{{r_o}}}$ and ${F_{{r_i}}}$ is described in Fig.~\ref{fig:Fig11}.

In evolution, the electric field induced by charged particles and particle free transport will expand the beam. Applied external electric field focalizes the beam. So there is a process in which expansion and contraction take place alternately. We describe evolution process of beam using electron density contour at different time step. The particle velocity is truncated at 6.0 with 64 points.

We simulated the evolution of a Gaussian beam and compared our results with the work of Ref.~\cite{filbet2001conservative}, which proves the accuracy that the UGKS obtains in plasma simulation.

Now, let's compare the computational cost with particle based methods introduced in Ref.~\cite{filbet2001conservative}. The test model is put in a $64 \times 64 \times 64 \times 64$ phase space.

We compared computation time of the UGKS and particle based scheme under the same phase space in Table~\ref{table:Table1}. The comparison results show a high efficiency achieved by the UGKS in solving the Vlasov equation. Although electron has wave-particle dualism, in this paper, we only study Gaussian beam under an electrodynamics frame. Based on the Vlasov-Poisson system, the motion of electrons is described in phase space with particle distribution function. Electrons are moving in space because of existence of electromagnet force. So the peak value of intensity of the beam varies with time in Fig.~\ref{fig:Fig12}. To obtain a correct evolution process of beam, the term of transport and acceleration in the Vlasov equation should be accurately coupled. Although the wave nature for electrons is out of the range in this paper, the Schr\"odinger equation can be used to construct a numerical method for quantum results of Gaussian beam case in the future.

\subsection{Linear Electron Plasma Wave Damping}
In this section, the effect of collisions on linear damping of spatially one-dimensional electron plasma wave (EPW) is simulated based on Boltzmann equation with electron-ion collision operator. The complete Boltzmann equation is written as \cite{banks2016vlasov}
\begin{equation}\label{Eq31}
\frac{{\partial f}}{{\partial t}} + {\bm u} \cdot \frac{{\partial f}}{{\partial {\bm x} }} + {\bm a}  \cdot \frac{{\partial f}}{{\partial {\bm u} }} =  - {C_{ei}}f,
\end{equation}
where the collision operator $ - {C_{ei}}f$ is given by
\begin{equation}\label{Eq32}
{C_{ei}}f =  - {\nu _{ei,th}}{\left| {{{{\bm u} }_{th}}} \right|^3}\frac{{\partial \left( {{\bm  P} } \right)}}{{\partial {\bm  u} }} \cdot \frac{{\partial f}}{{\partial {\bm u} }},
\end{equation}
where ${\nu _{ei,th}}$ denotes the thermal electron-ion collision frequency and ${\bm u} _{th}$ is velocity of particle thermal motion. Tensor $\bm P$ is defined by
\begin{equation}\label{Eq33}
{\bm P}  = \frac{1}{{{{{\bm u} }^3}}}\left( {{{{\bm u} }^2}{\bm I}  - {\bm u} :{\bm u} } \right).
\end{equation}

In the EPW case, the wavenumber is set to be 0.3. In the low perturbation amplitude regime, the initial electron distribution is
\begin{equation}\label{Eq34}
f = \rho \left( {\frac{\lambda }{\pi }} \right){e^{ - \lambda \left( {{{{\bm u} }^2}} \right)}}\left( {1 + \alpha \cos \left( {kx} \right)} \right),
\end{equation}
where $\alpha  = 0.0001$ and $k = \frac{\pi }{{250}}$. The 1D computational domain is discretized with 64 grid points. In microscopic velocity space, 128 discrete points are used. To compared with the result of Ref.~\cite{banks2016vlasov}, the thermal electron-ion collision frequency ${\nu _{ei,th}}$ is set to be 0.05. The amplitude of the perturbation's electric field E is plotted on a log scale as a function of time in Fig.~\ref{fig:Fig13}.

Where ${\omega _{pe}}$ is the circular frequency of initial perturbation. The result obtained from present methods is consistent with Ref.~\cite{banks2016vlasov}. The decaying process fit well with exponential decaying line. Because of the initial perturbation of particle distribution, the electric field is also damped in oscillatory process. In traditional numerical method based on continuous assumption, it¡¯s very hard to capture these phenomena for the lack of direct modeling of particle motion. For particle based method, the huge number of particles always produces unacceptable computational cost. In UGKS, particle motion is described in modeling of flux through cell interface under discrete microscopic velocity space. For every discrete distribution function, computation cost only comes from process of FVM. But accuracy is the same as particle based method since all particles are considered based on statistical concept using distribution function. With suitable collision model, different state of plasma can be accurately obtained.

\section{Conclusions}\label{sec:Conclusions}
In this paper, we solve the Vlasov equation based on the UGKS. This approach gives a full representation for particle moving in phase space with long-range electromagnet interaction. An integral solution of the kinetic model under discrete physical and particle velocity space is applied. In non-equilibrium state, this method can describe distribution function with discrete phase space. Compared to Lagrangian method, the UGKS has a much lower computational cost but achieves the same accuracy in plasma simulation. Many methods for solution of the Vlasov equation, such as FEM and spectral method, still have to solve equation of particle motion to decide the coefficients in basis functions for particle velocity dimensions. As a result, the time step is restricted according to dissipative length scale determined by the physical transport. Time step in current work can be decided by CFL strategy without spurious dissipation. In order to simplify the solution of the Poisson's equation by using the Green's function, we restrict the boundary conditions to the periodic or symmetric boundary conditions. The method presented in this paper can be used in simulation for flows of charged particles in plasma. In multi-scale problems of plasma flows from free transport regime to collisional effect, UGKS can give pretty good description of physical evolution. Compared to particle based method, computational efficiency has also been greatly improved. From the results via the UGKS, it is proved to be a better way than previous particle methods to study plasma problems because of its good accuracy and low computational cost.

\section*{Acknowledgements}
This research project was financially supported by the National Natural Science Foundation of China (Grant No. 11472219), the Natural Science Basic Research Plan in Shaanxi Province of China (Program No. 2015JM1002), the Fundamental Research Funds for the Central Universities (A020415), as well as the 111 Project of China (B17037).

\section*{References}
\bibliographystyle{elsarticle-num}
\bibliography{plasma_UGKS}




\clearpage
\renewcommand\thefigure{\arabic{table}}

\begin{table}
\caption{Computation time for comparison under $64 \times 64 \times 64 \times 64$ phase space.}\label{table:Table1}
\centering
\begin{tabular}{c|c|c}
\hline
\hline
\hspace{4mm}Number of processors \hspace{4mm} & \hspace{4mm}Unified Gas Kinetic Scheme \hspace{4mm}& \hspace{4mm}Particle based method \hspace{4mm}\\
\hline
 2 &  1 h 18 min & 6 h 20 min \\
\hline
 4  &  48.63min  & 2 h 11 min \\
\hline
\hline
\end{tabular}
\end{table}

\clearpage
\renewcommand\thefigure{\arabic{figure}}

\begin{figure}
  \centering
  \includegraphics[width=0.5\textwidth]{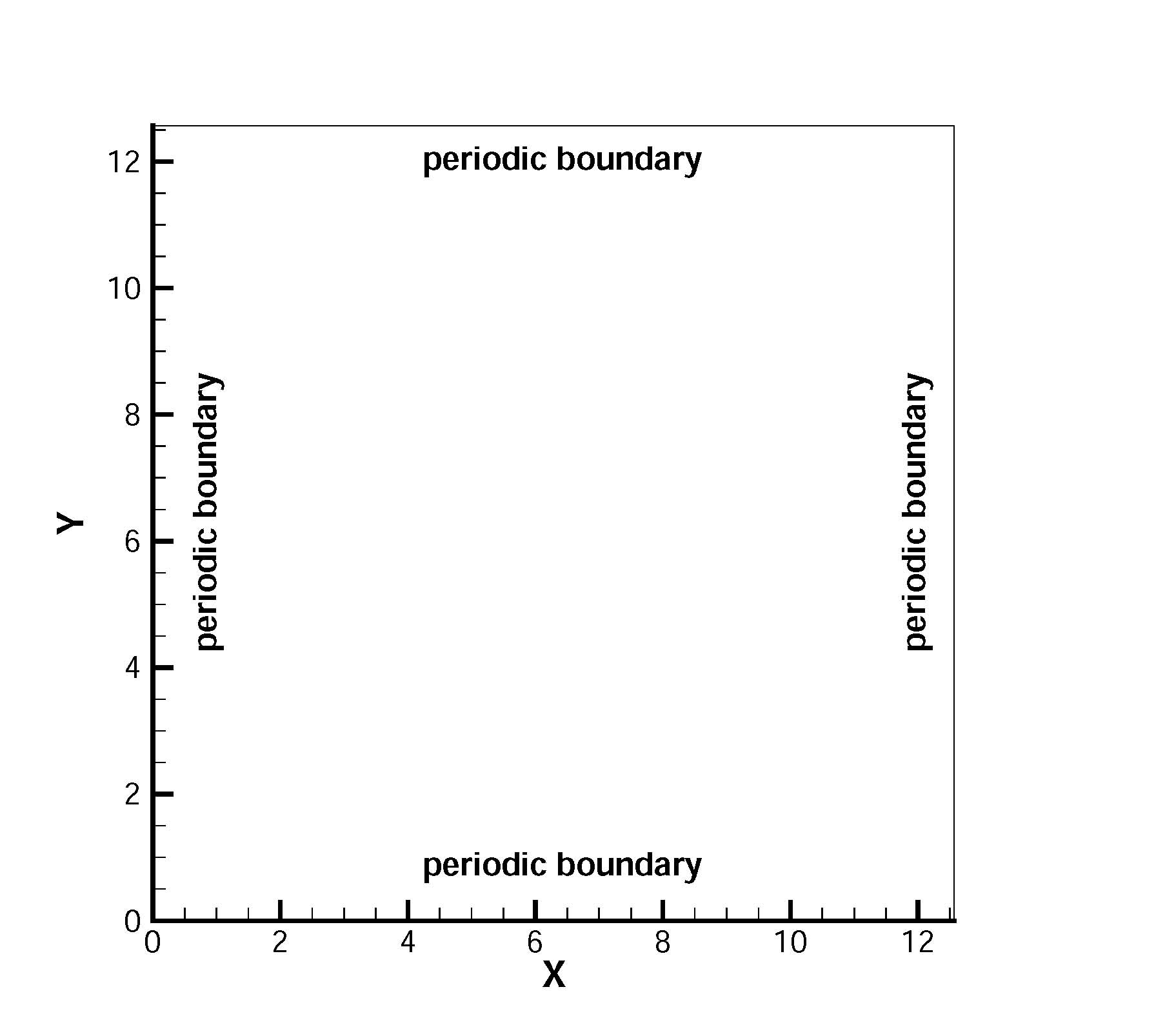}
  \caption{Boundary condition for the simulation of the linear Landau damping.}
  \label{fig:Fig01}
\end{figure}

\begin{figure}
  \centering
  \includegraphics[width=0.5\textwidth]{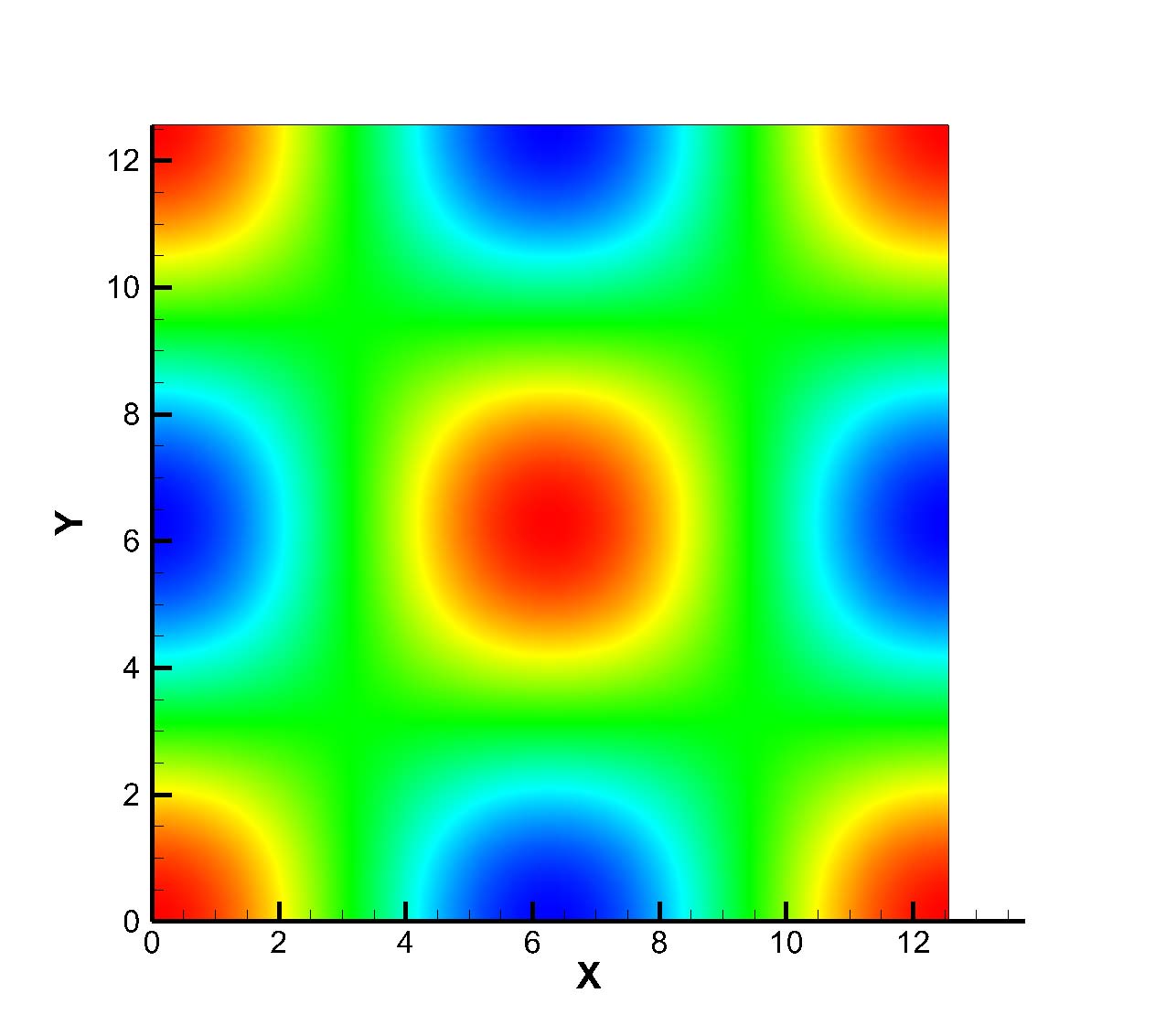}
  \caption{Initial electron density contour of the linear Landau damping.}
  \label{fig:Fig02}
\end{figure}

\begin{figure}
  \centering
  \subfigure[$ t = 0.5$]{\label{fig:Fig03a}\includegraphics[width=0.45\textwidth]{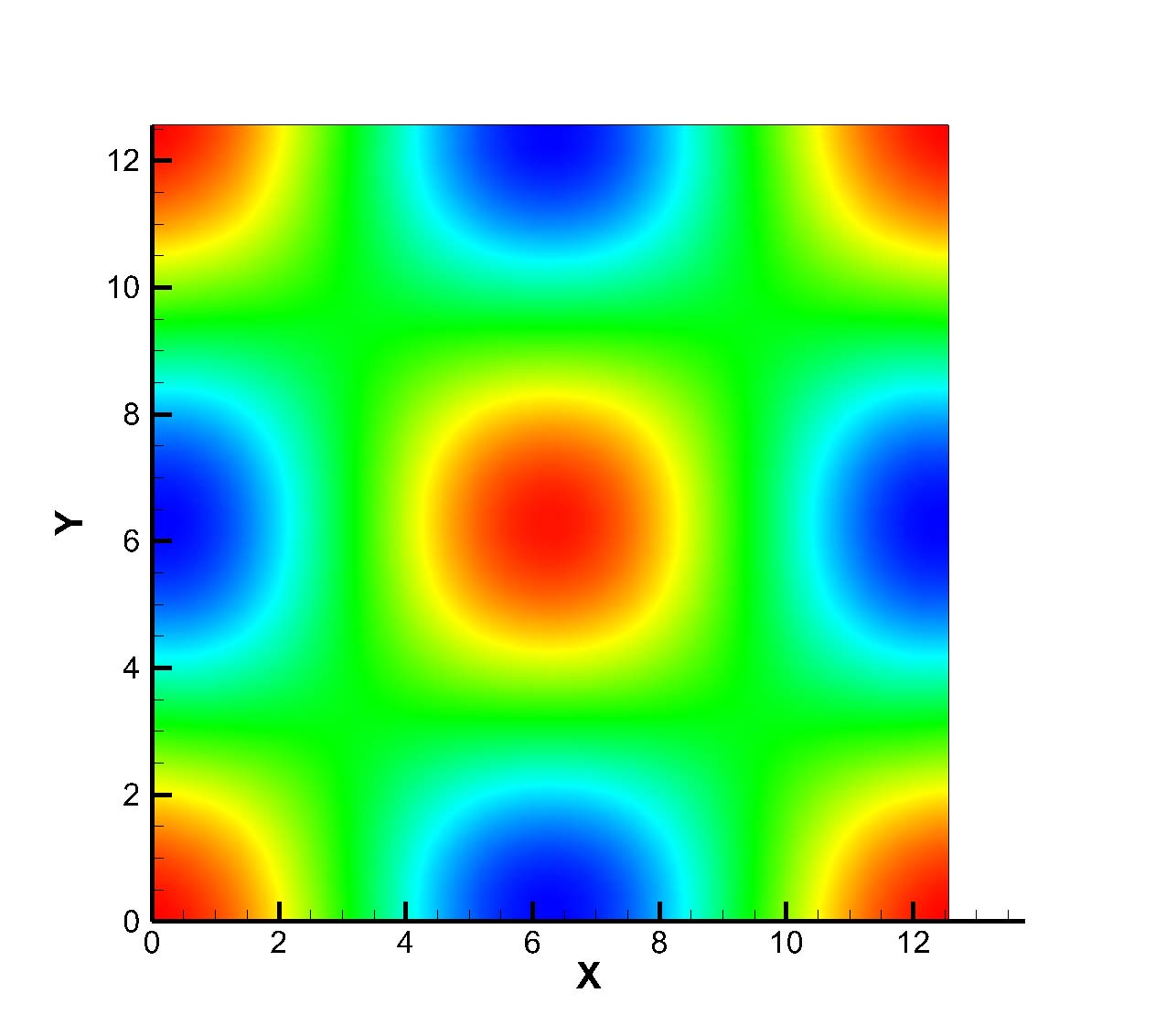}}
  \subfigure[$ t = 1.0$]{\label{fig:Fig03b}\includegraphics[width=0.45\textwidth]{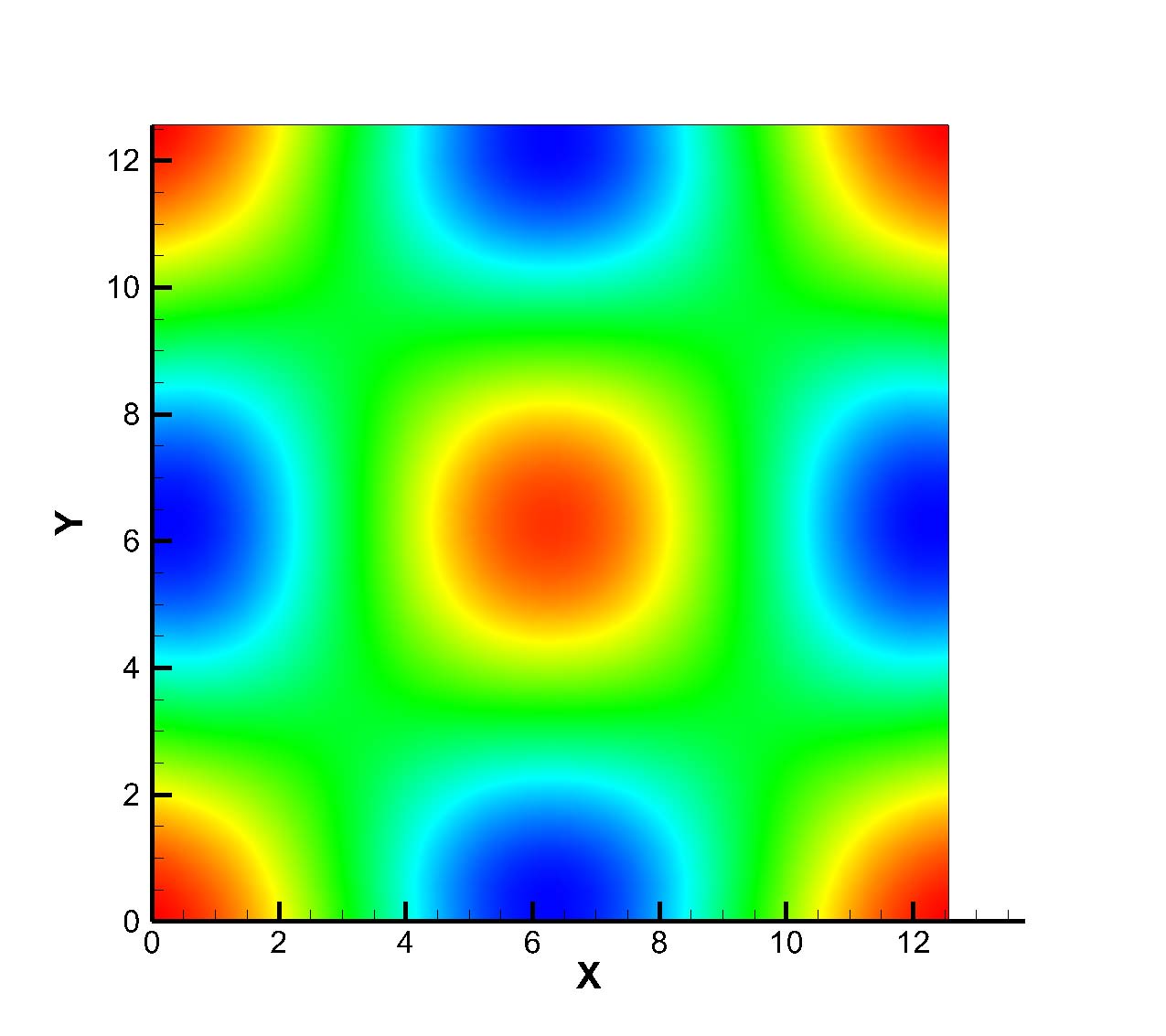}}
  \subfigure[$ t = 1.5$]{\label{fig:Fig03c}\includegraphics[width=0.45\textwidth]{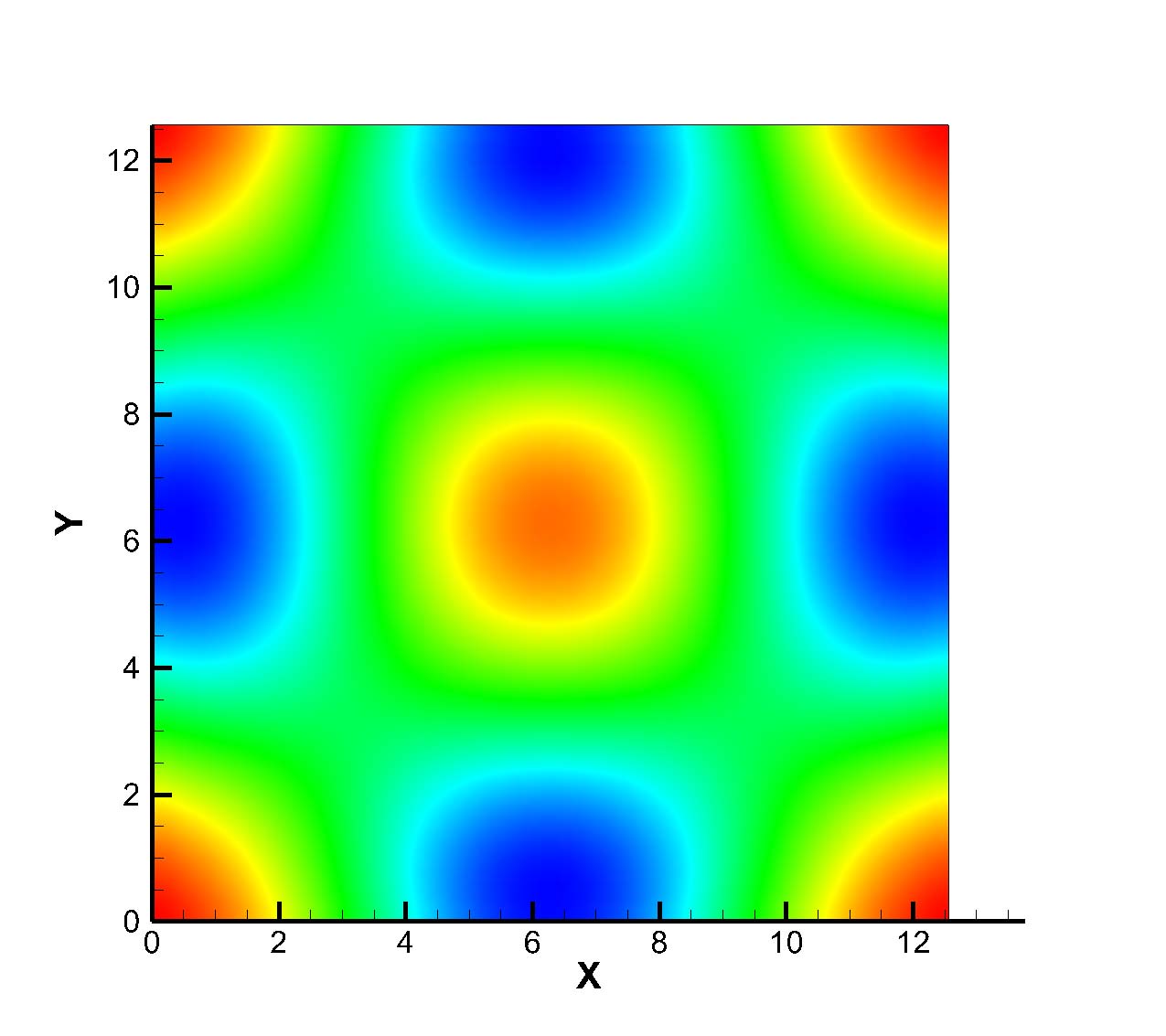}}
  \subfigure[$ t = 2.0$]{\label{fig:Fig03d}\includegraphics[width=0.45\textwidth]{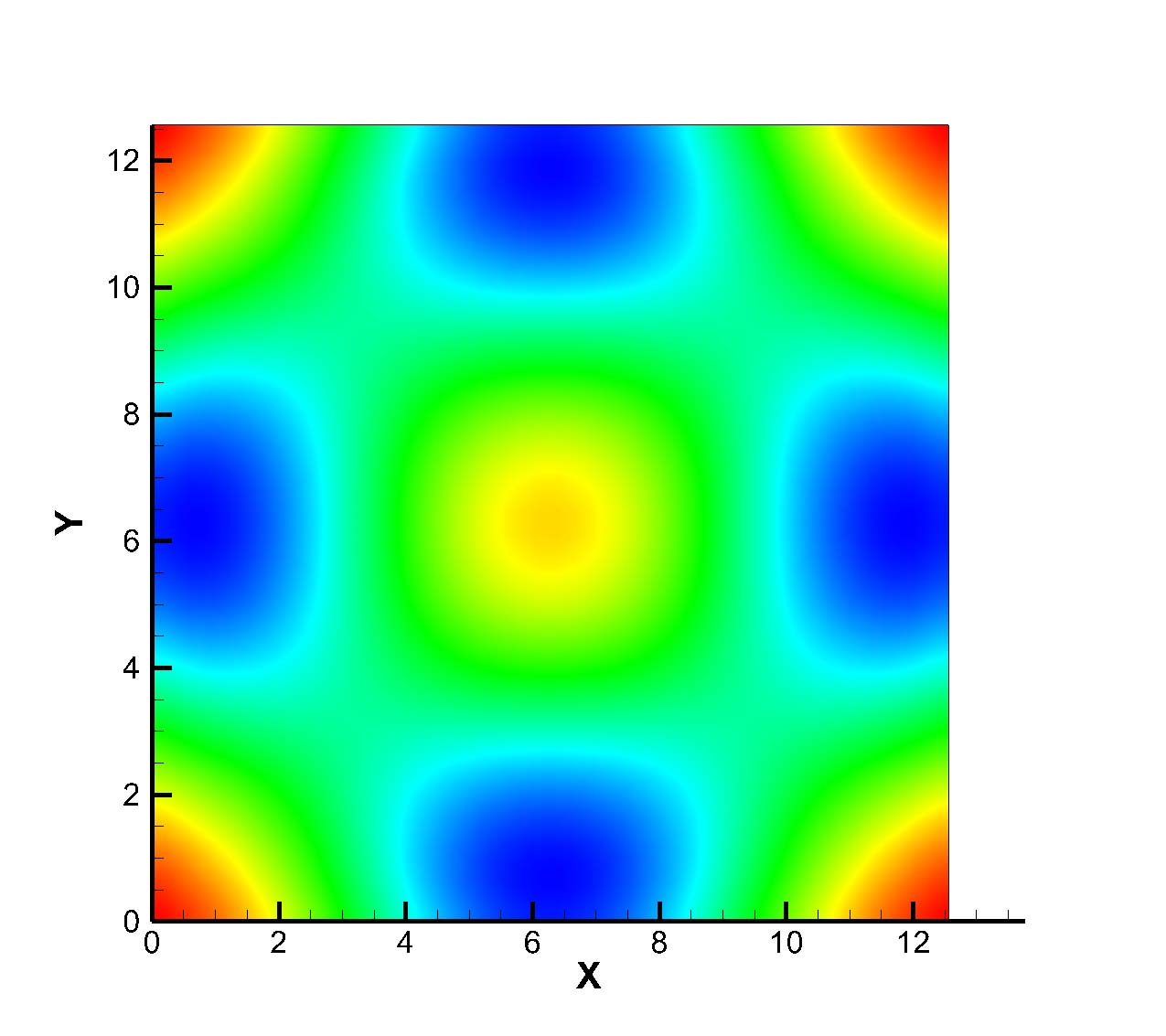}}
  \subfigure[$ t = 2.5$]{\label{fig:Fig03e}\includegraphics[width=0.45\textwidth]{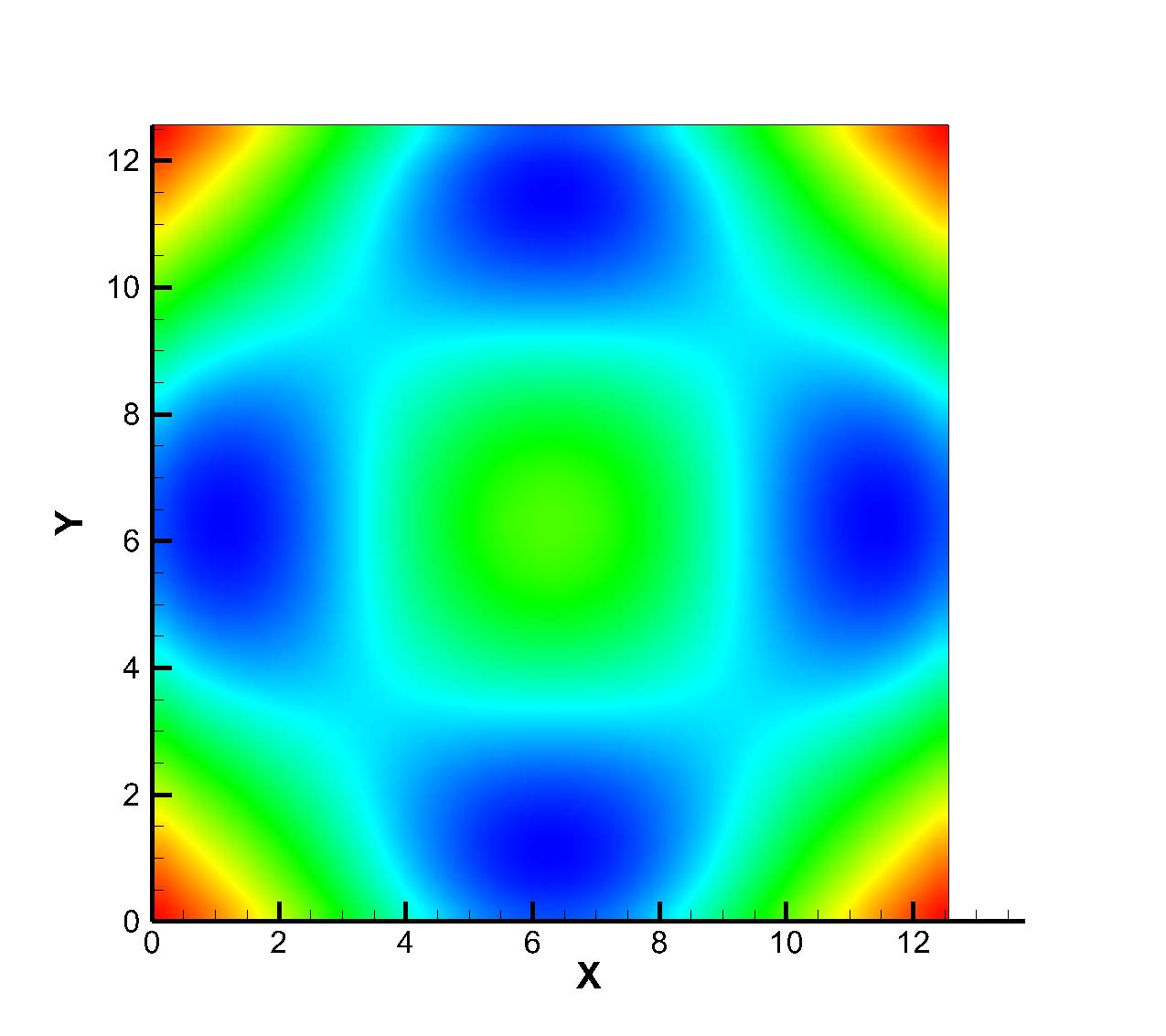}}
  \subfigure[$ t = 3.0$]{\label{fig:Fig03f}\includegraphics[width=0.45\textwidth]{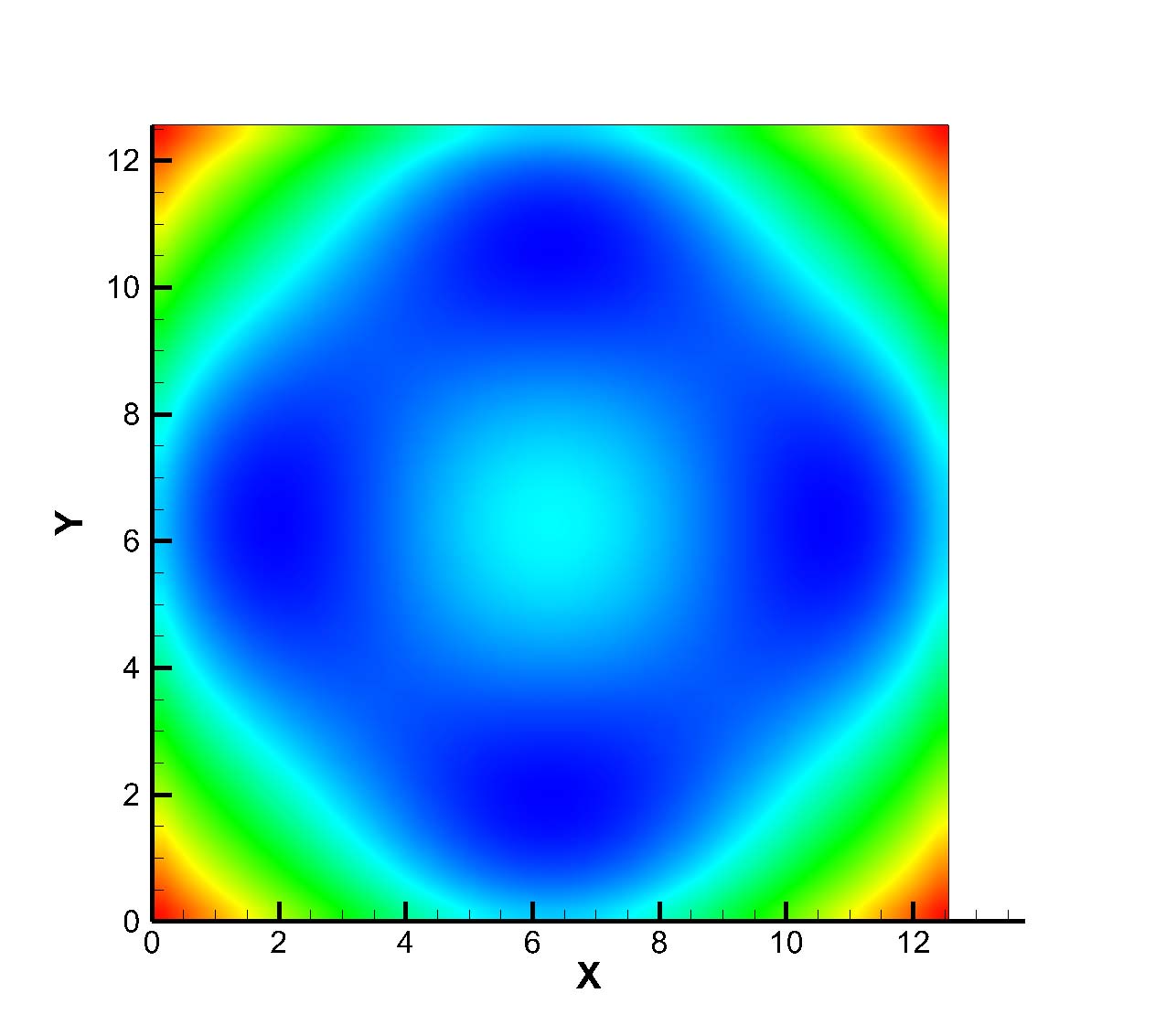}}
  \caption{Density contour of the linear Landau damping at different time.}
  \label{fig:Fig03}
\end{figure}

\begin{figure}
  \centering
  \includegraphics[width=0.5\textwidth]{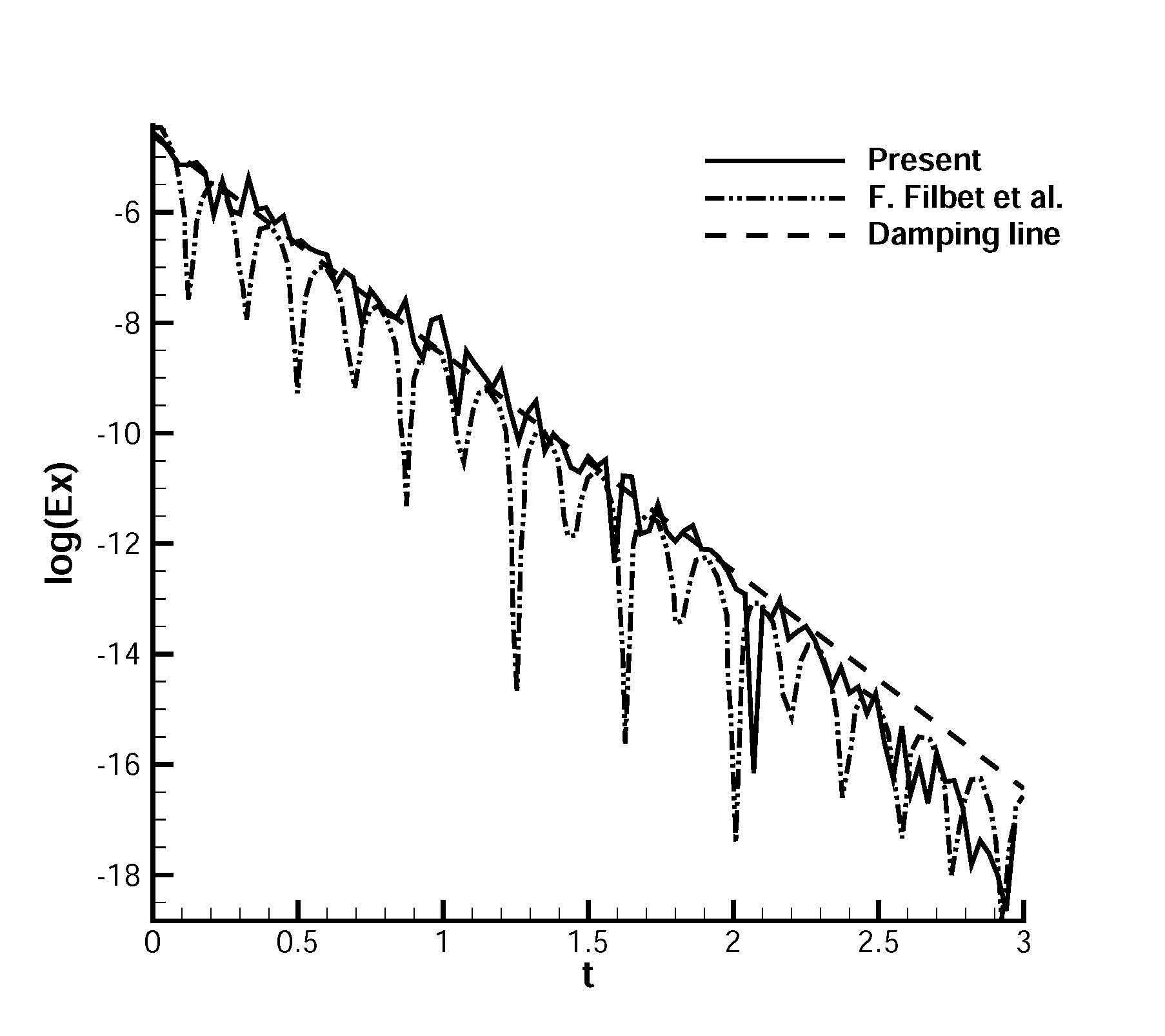}
  \caption{Evolution of symmetric electric field of the linear Landau damping.}
  \label{fig:Fig04}
\end{figure}

\begin{figure}
  \centering
  \includegraphics[width=0.5\textwidth]{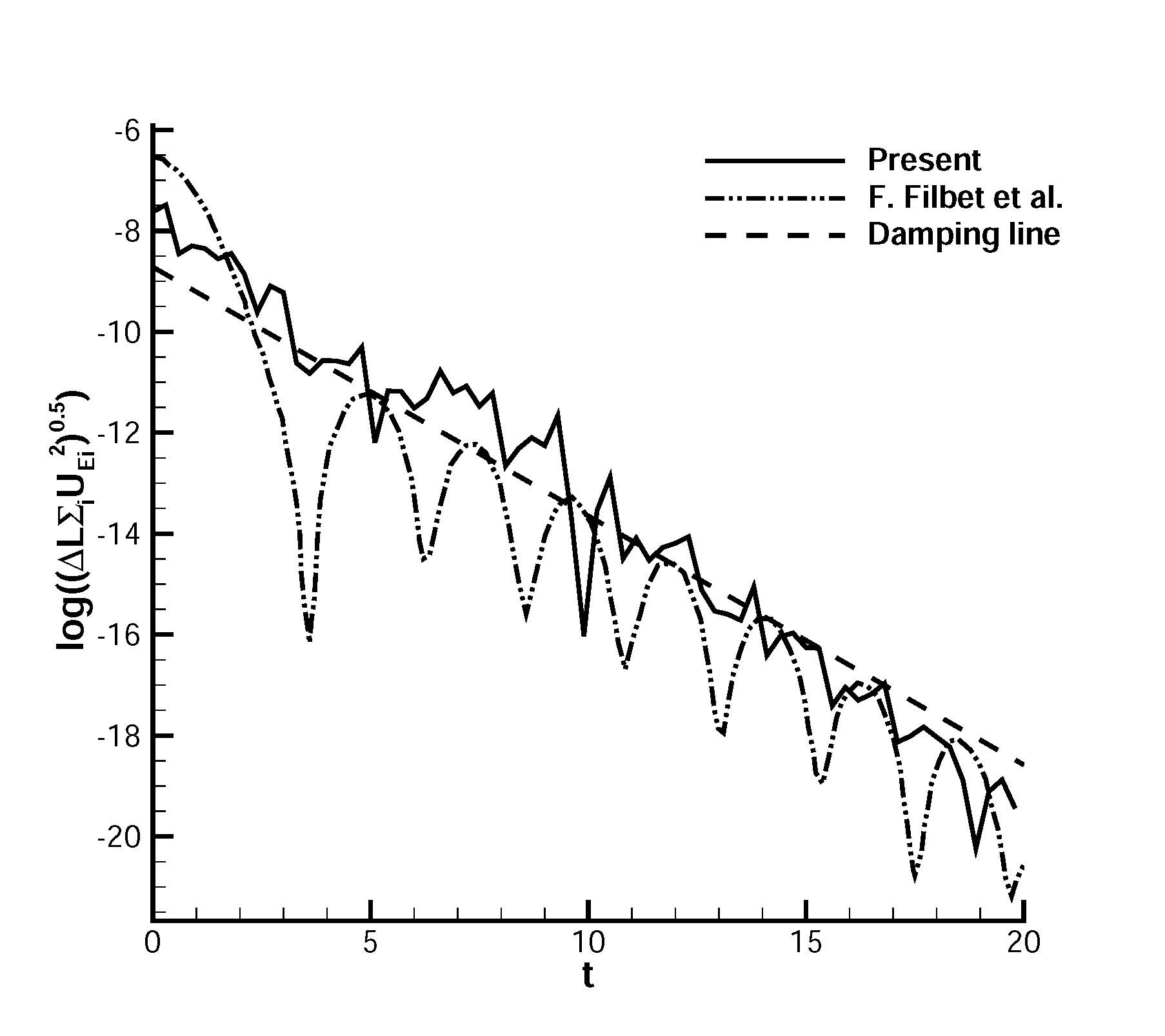}
  \caption{Evolution of electric potential energy of the linear Landau damping.}
  \label{fig:Fig05}
\end{figure}

\begin{figure}
  \centering
  \includegraphics[width=0.5\textwidth]{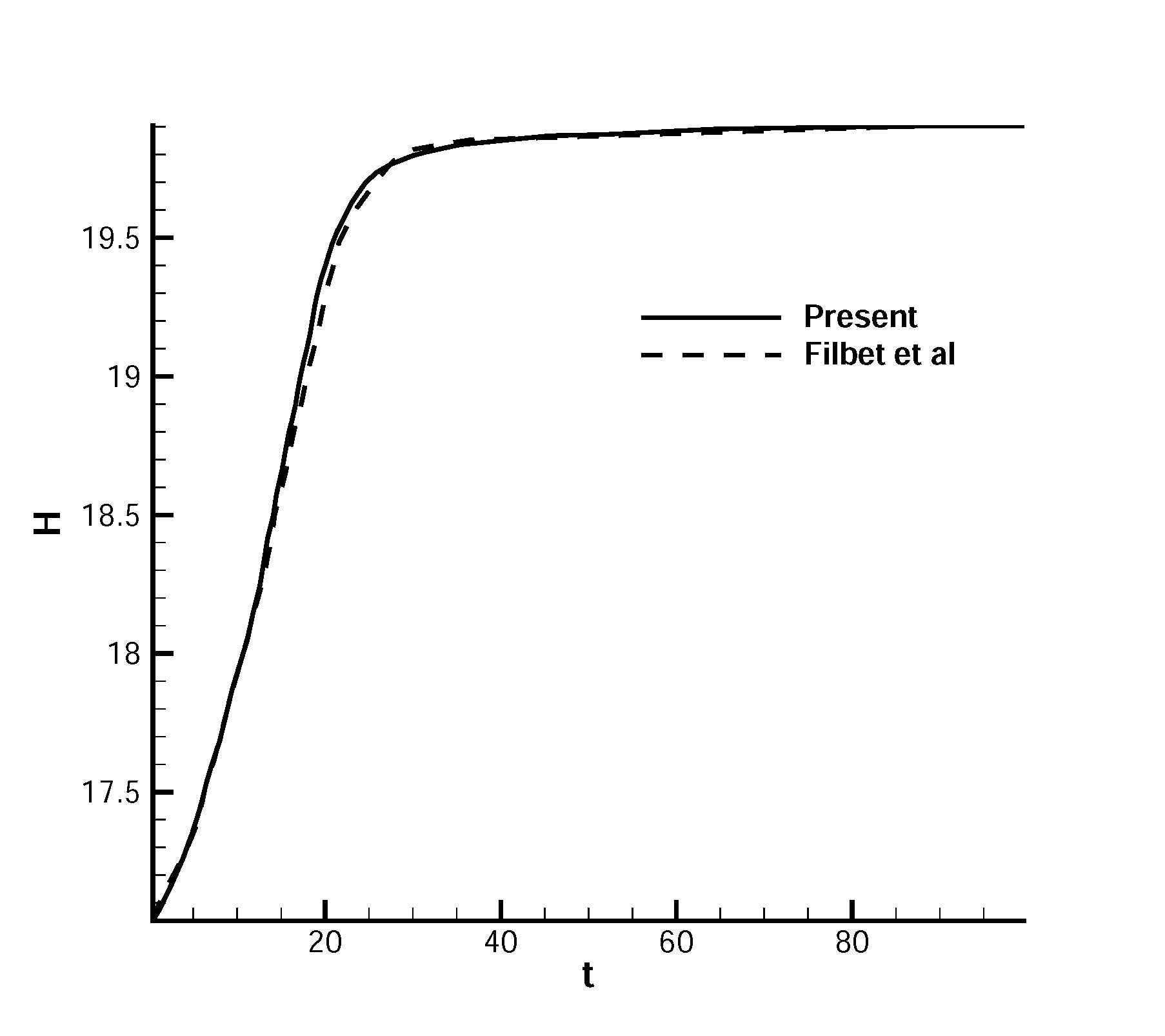}
  \caption{Evolution of the kinetic entropy of nonlinear Landau damping.}
  \label{fig:Fig06}
\end{figure}

\begin{figure}
  \centering
  \includegraphics[width=0.5\textwidth]{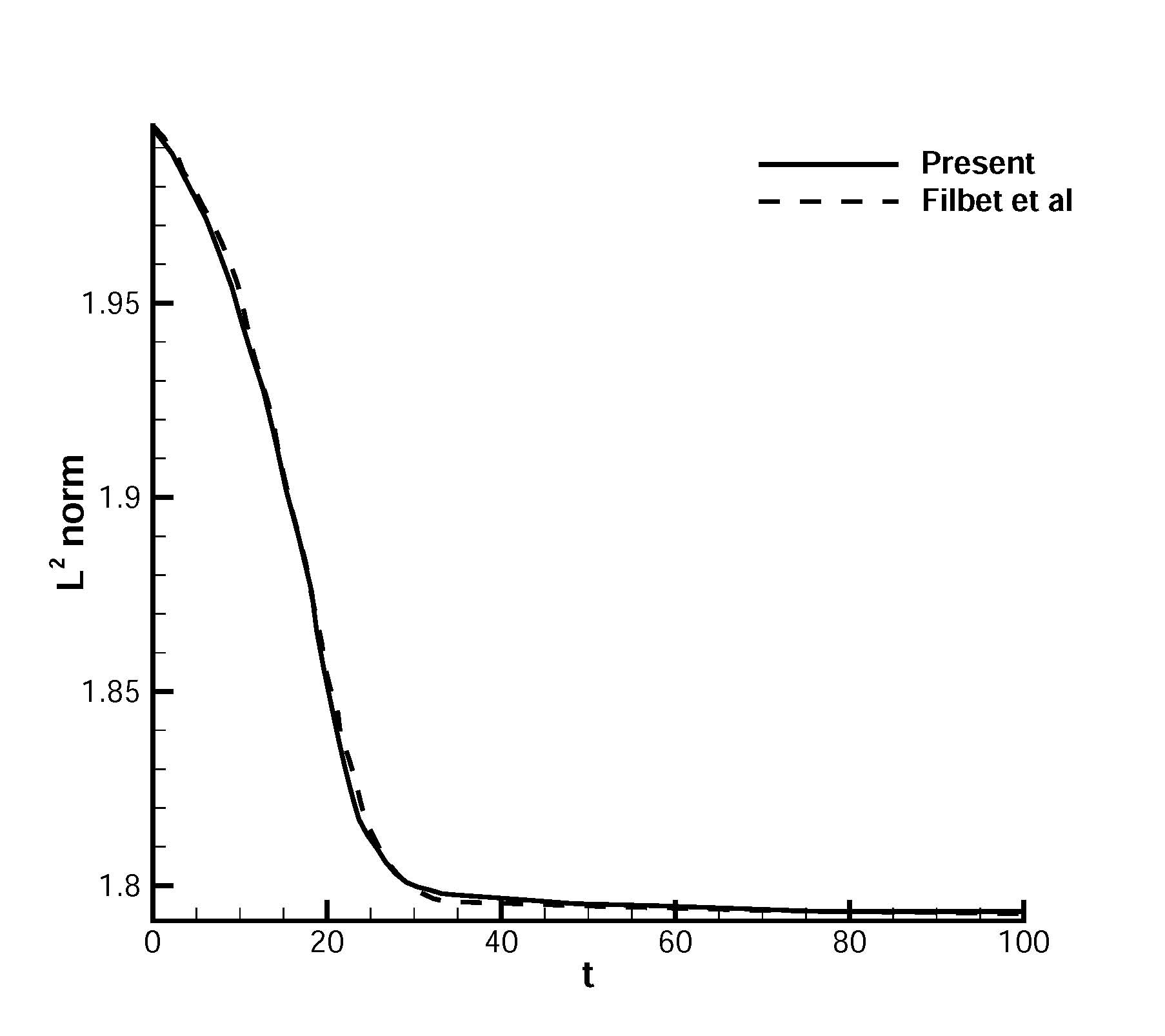}
  \caption{Evolution of $L^2 - norm$ of distribution function in nonlinear Landau damping.}
  \label{fig:Fig07}
\end{figure}

\begin{figure}
  \centering
  \subfigure[$ t = 5$]{\label{fig:Fig08a}\includegraphics[width=0.45\textwidth]{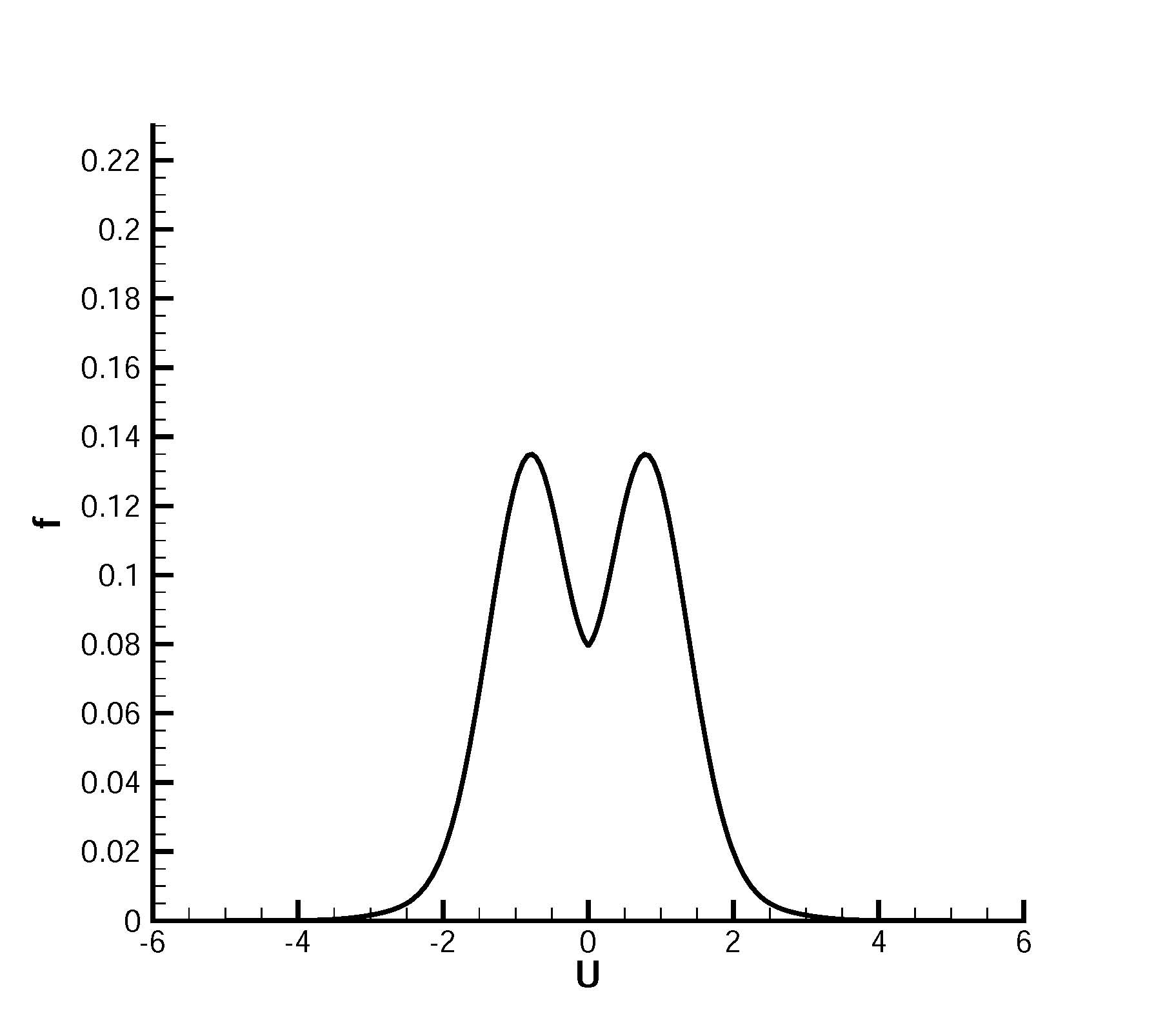}}
  \subfigure[$ t = 10$]{\label{fig:Fig08b}\includegraphics[width=0.45\textwidth]{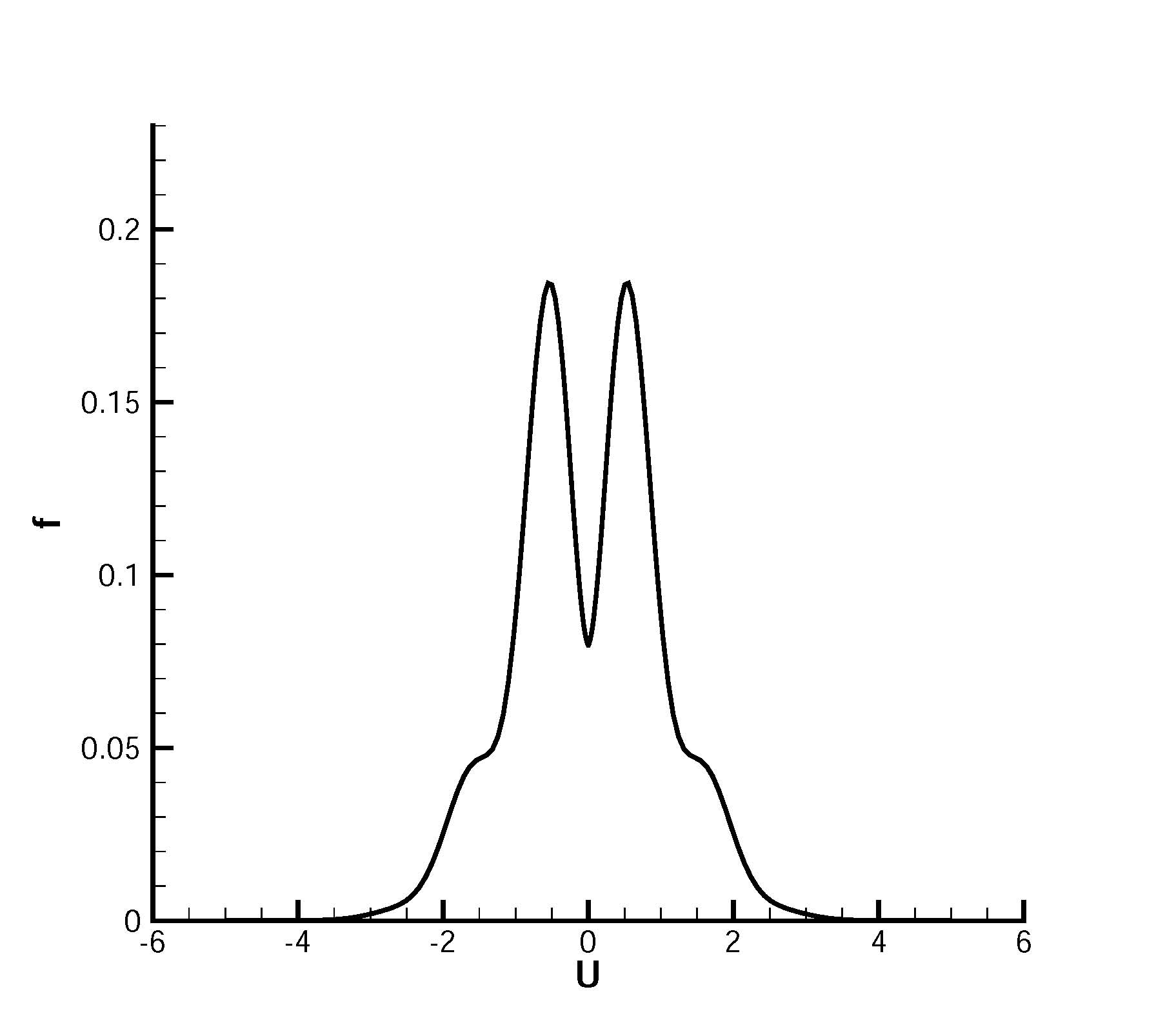}}
  \subfigure[$ t = 15$]{\label{fig:Fig08c}\includegraphics[width=0.45\textwidth]{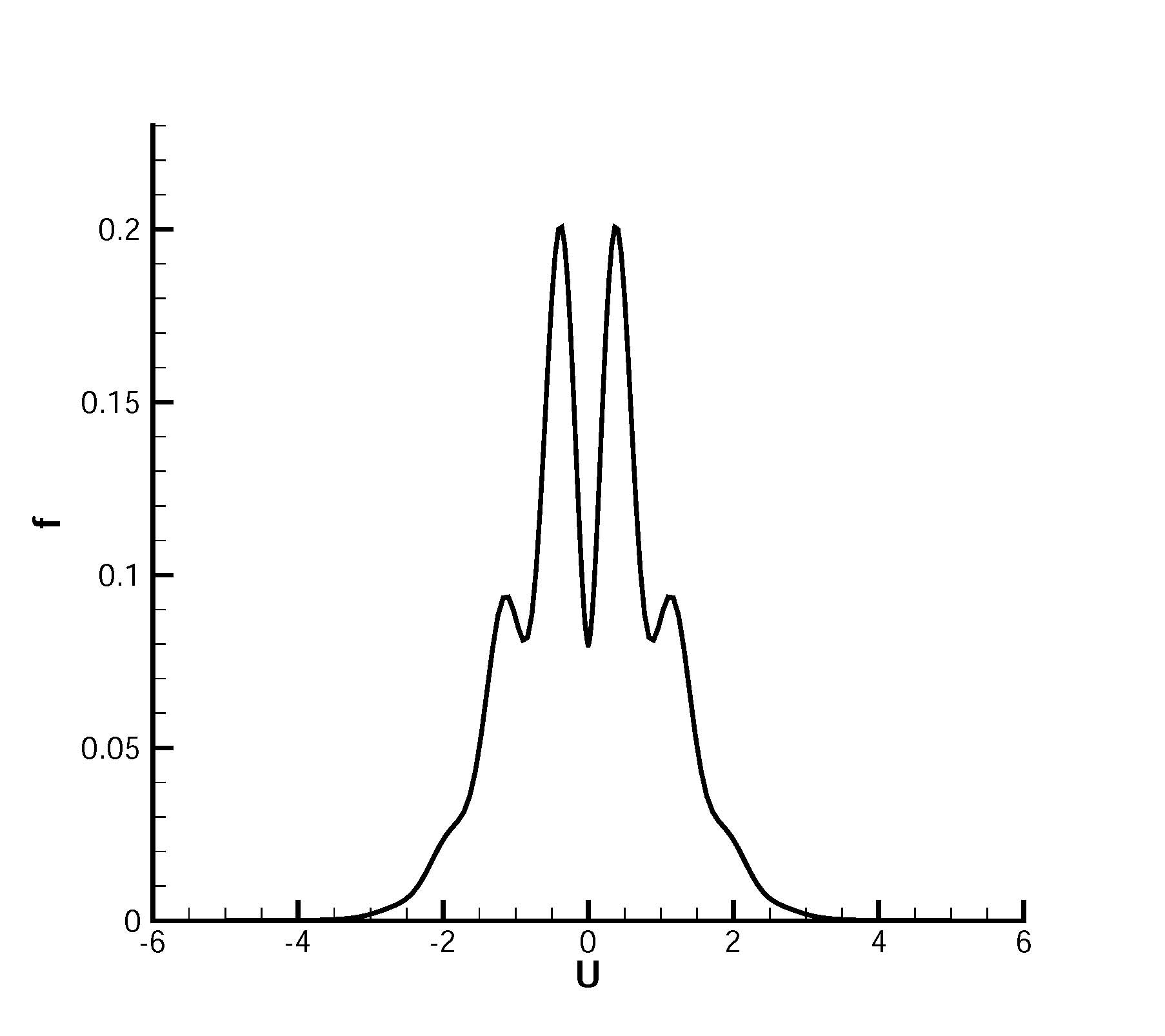}}
  \subfigure[$ t = 20$]{\label{fig:Fig08d}\includegraphics[width=0.45\textwidth]{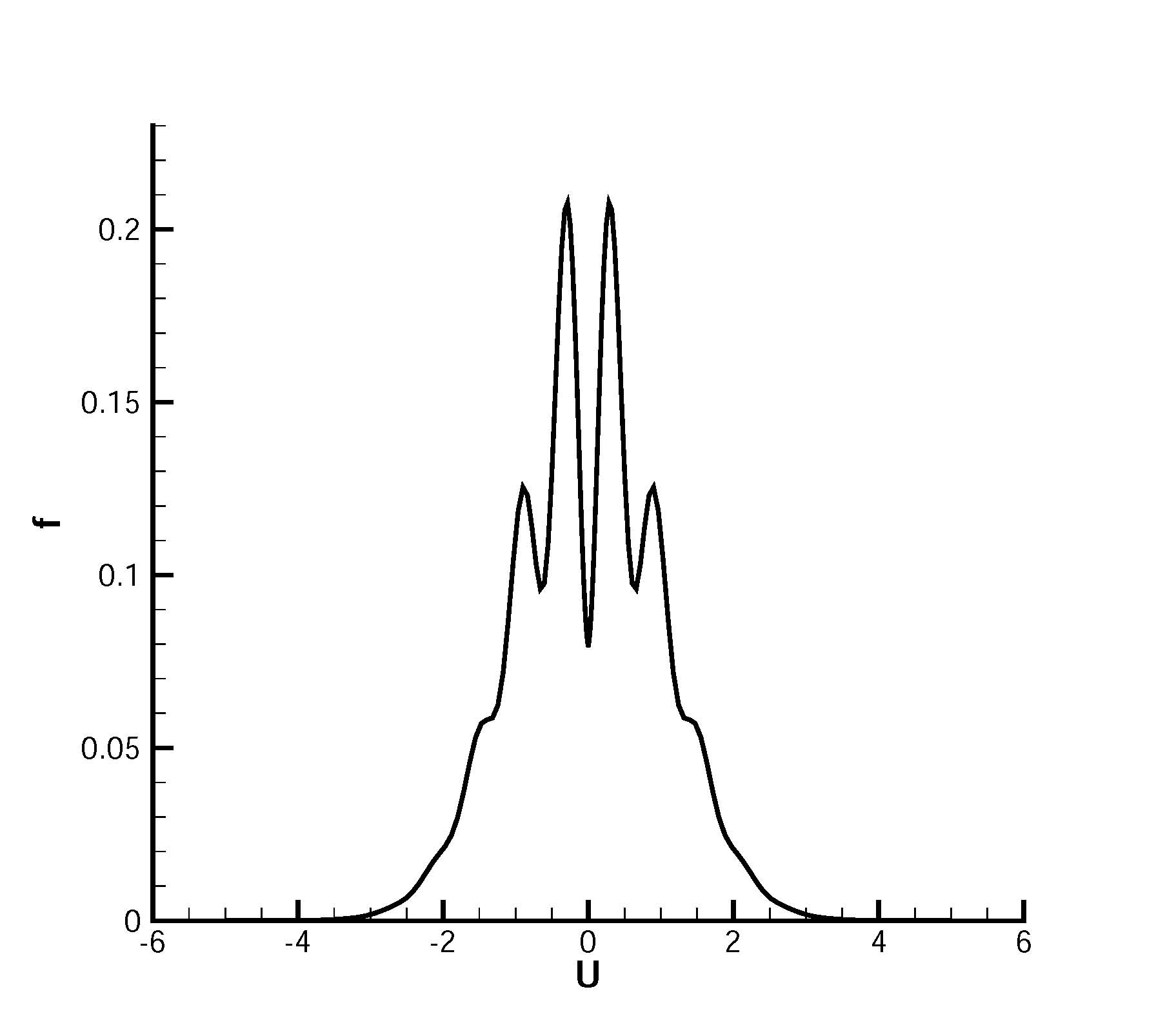}}
  \subfigure[$ t = 25$]{\label{fig:Fig08e}\includegraphics[width=0.45\textwidth]{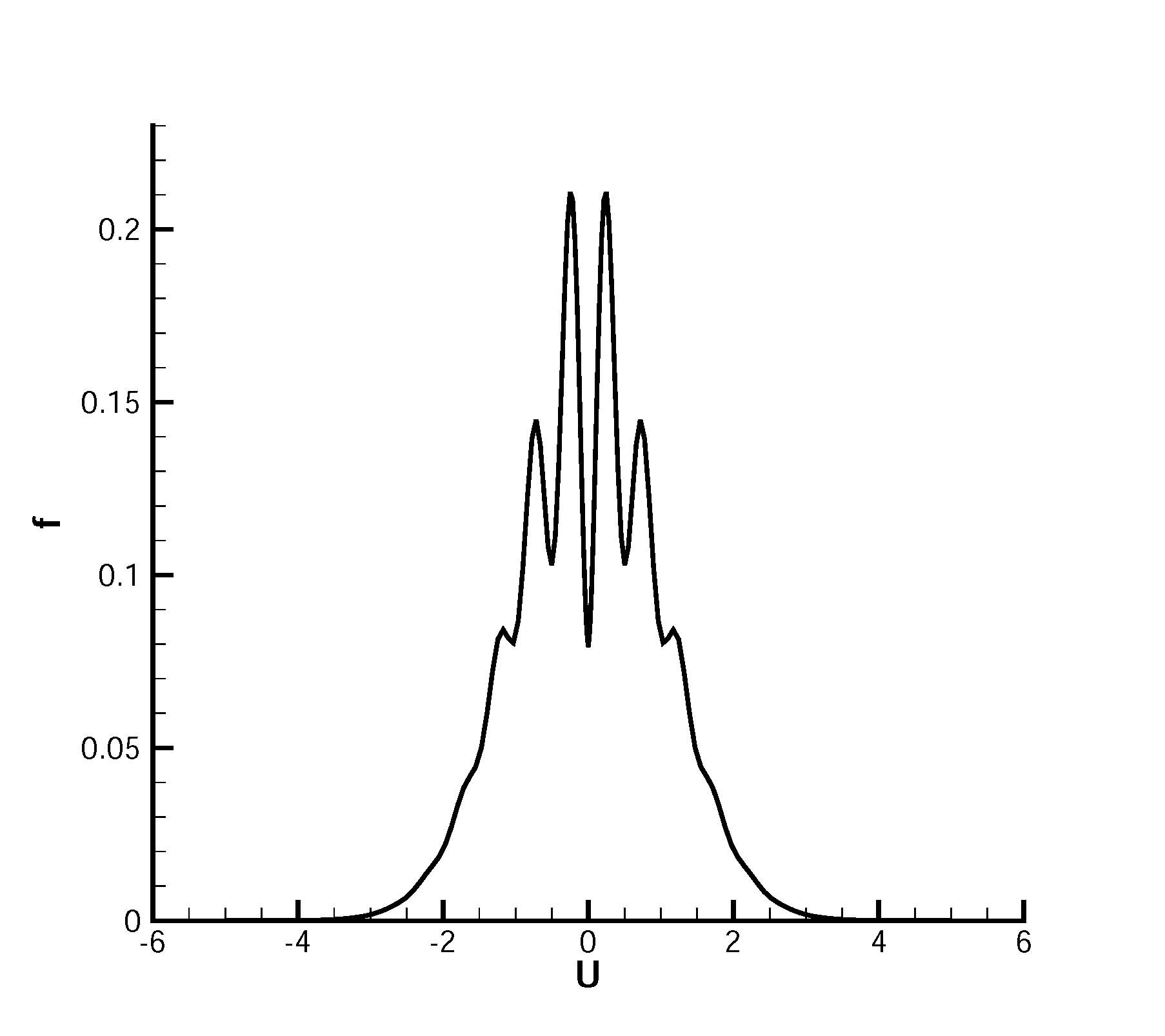}}
  \subfigure[$ t = 30$]{\label{fig:Fig08f}\includegraphics[width=0.45\textwidth]{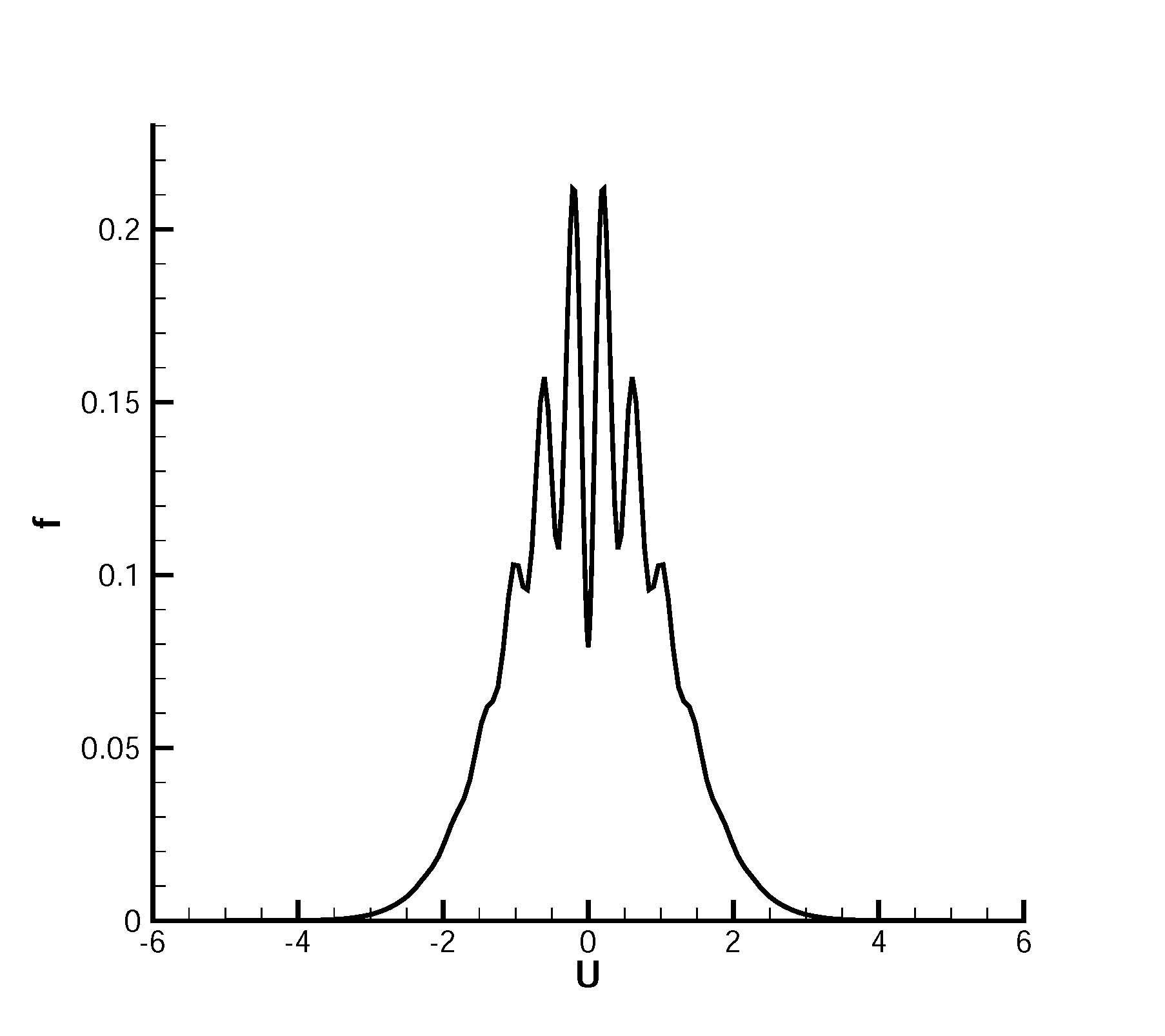}}
  \caption{Non-equlibrium distribution functions at different time steps of nonlinear Landau damping.}
  \label{fig:Fig08}
\end{figure}

\begin{figure}
  \centering
  \subfigure[$ t = 5$]{\label{fig:Fig09a}\includegraphics[width=0.45\textwidth]{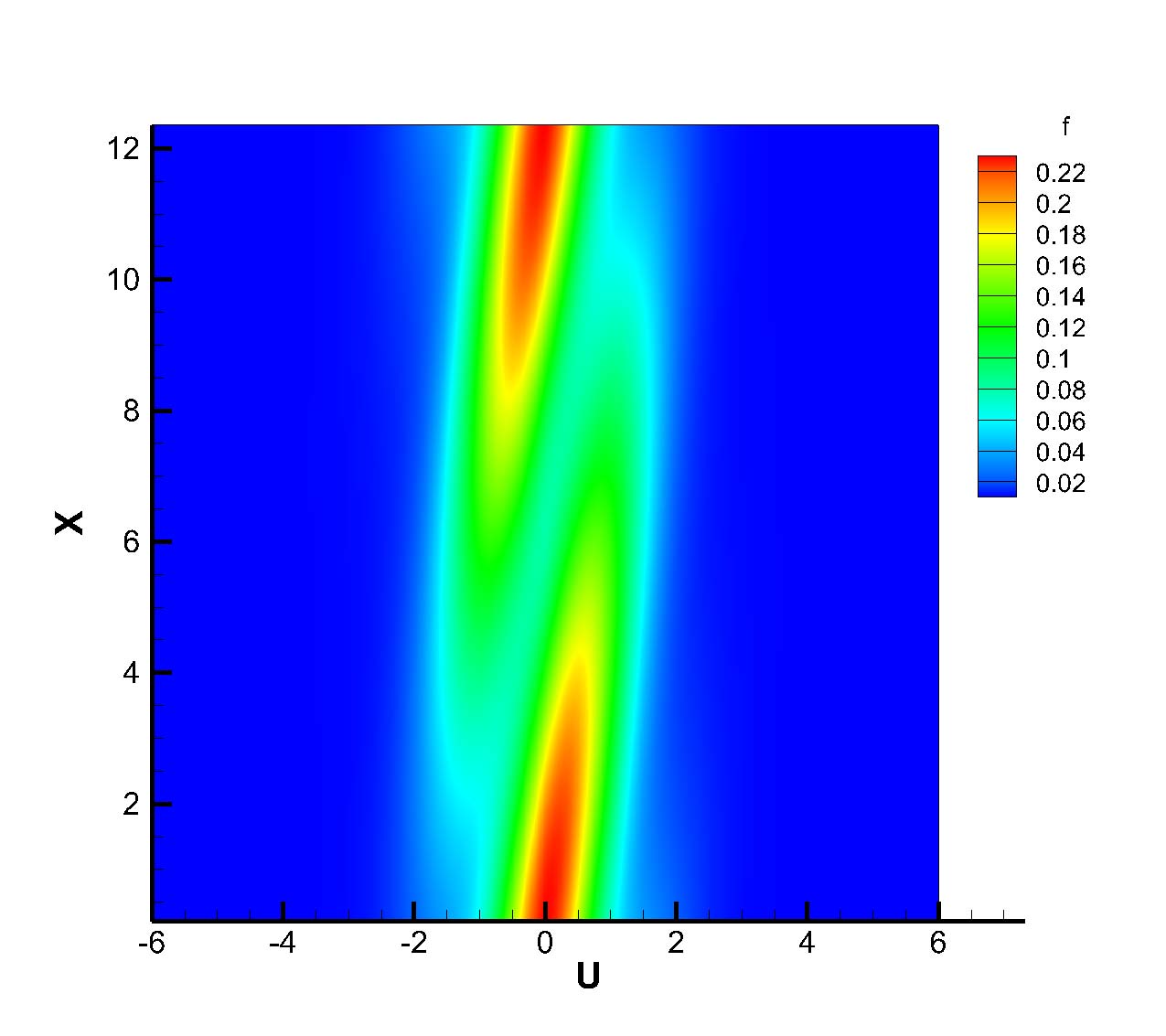}}
  \subfigure[$ t = 10$]{\label{fig:Fig09b}\includegraphics[width=0.45\textwidth]{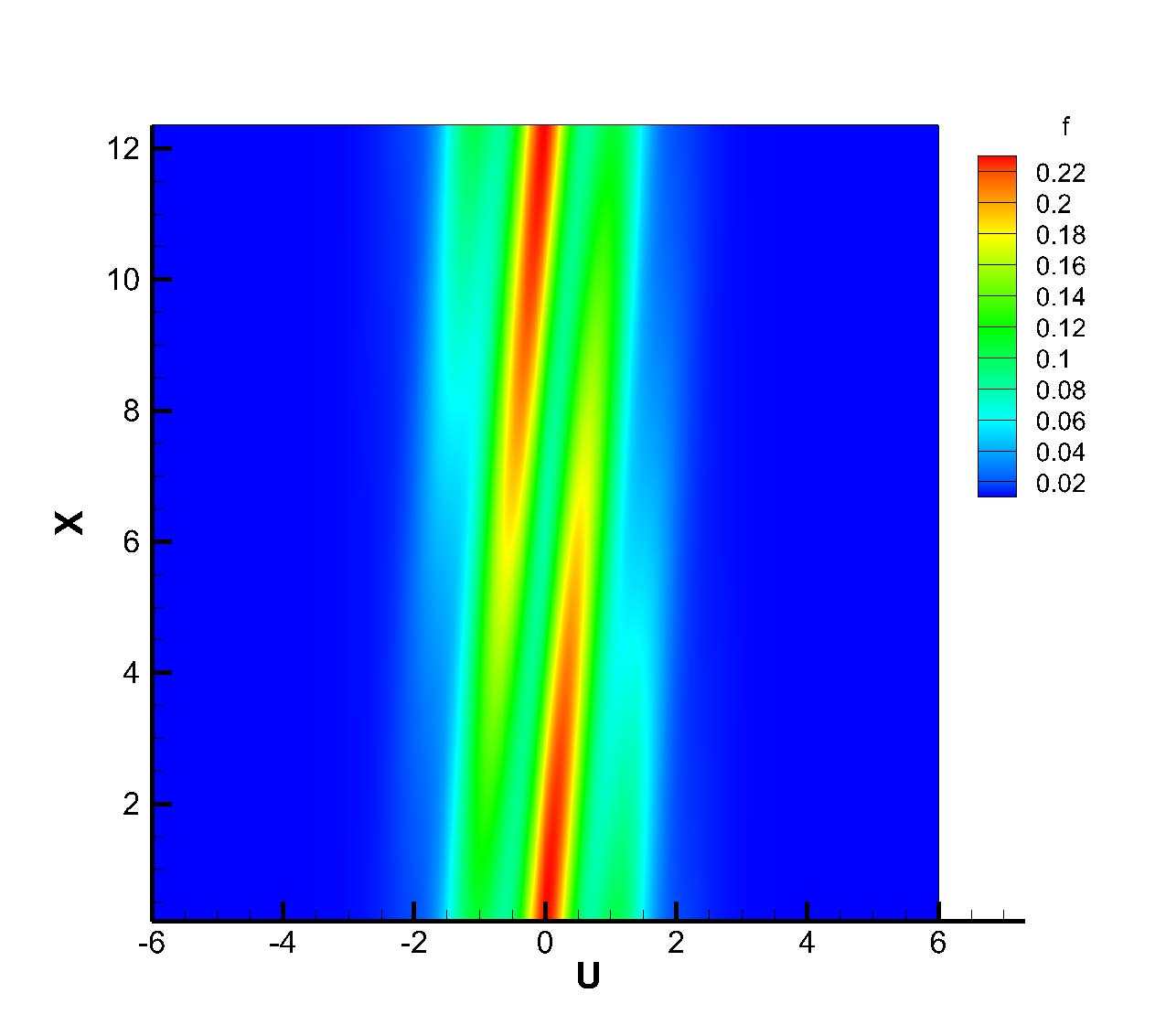}}
  \subfigure[$ t = 15$]{\label{fig:Fig09c}\includegraphics[width=0.45\textwidth]{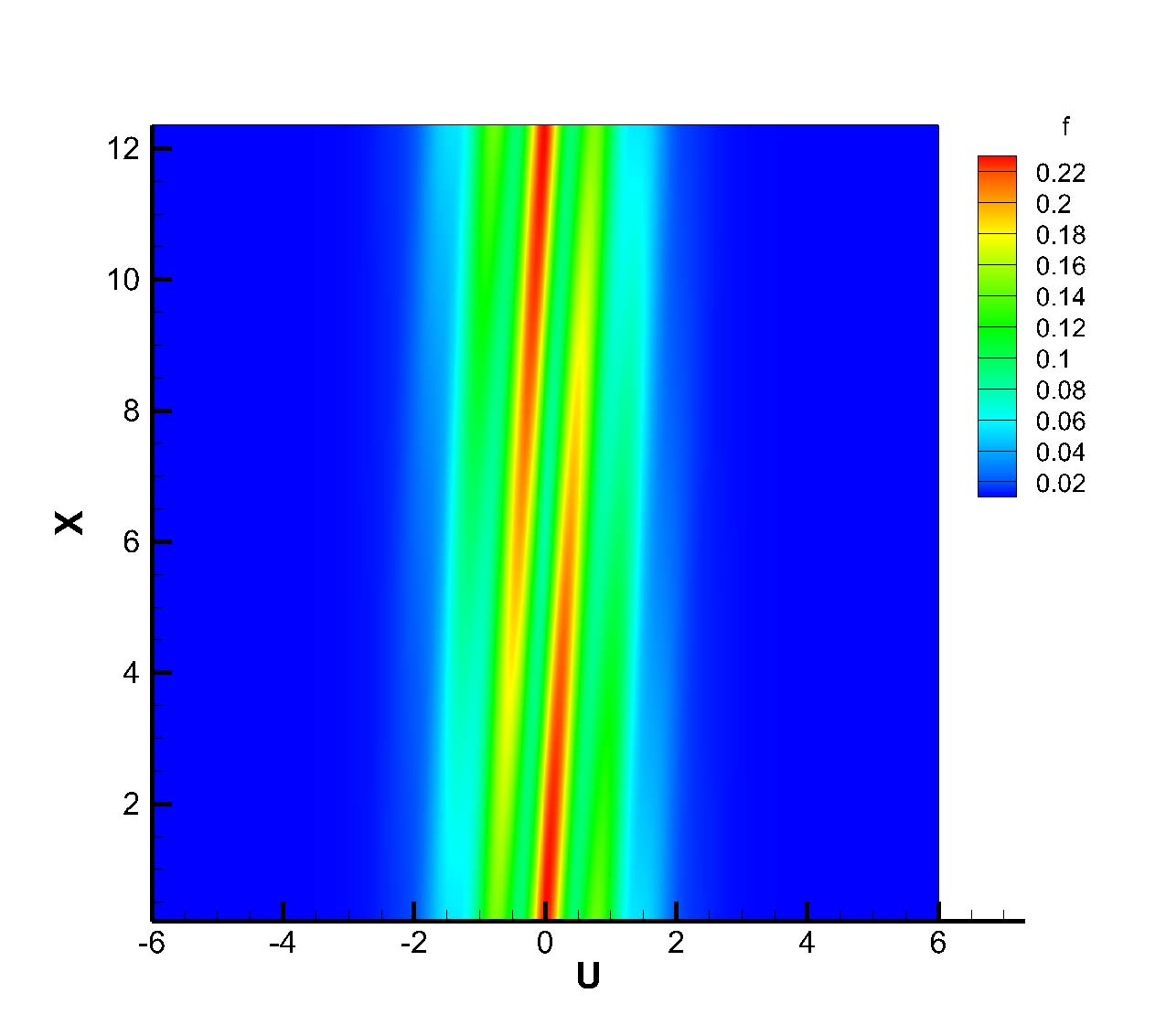}}
  \subfigure[$ t = 20$]{\label{fig:Fig09d}\includegraphics[width=0.45\textwidth]{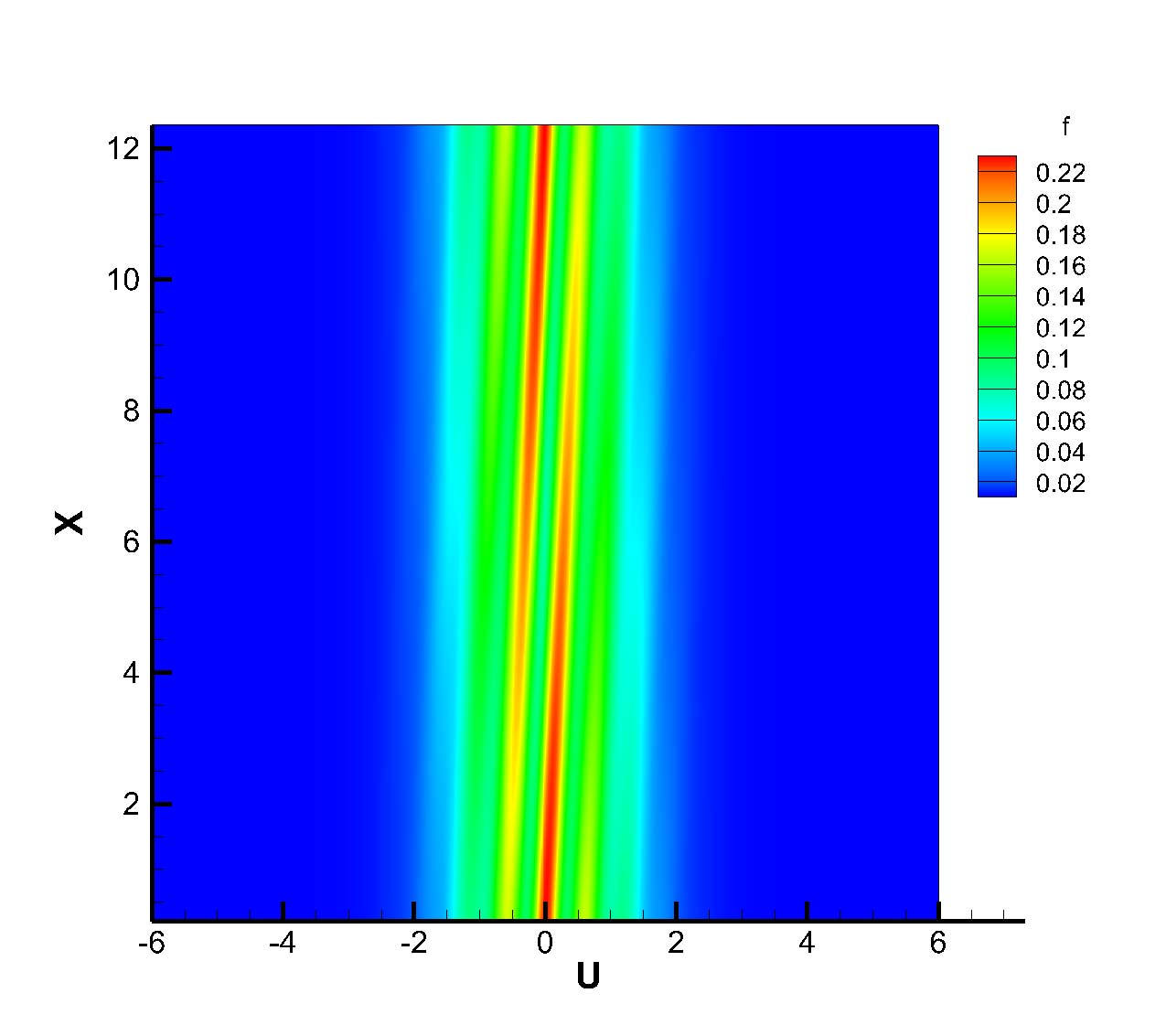}}
  \subfigure[$ t = 25$]{\label{fig:Fig09e}\includegraphics[width=0.45\textwidth]{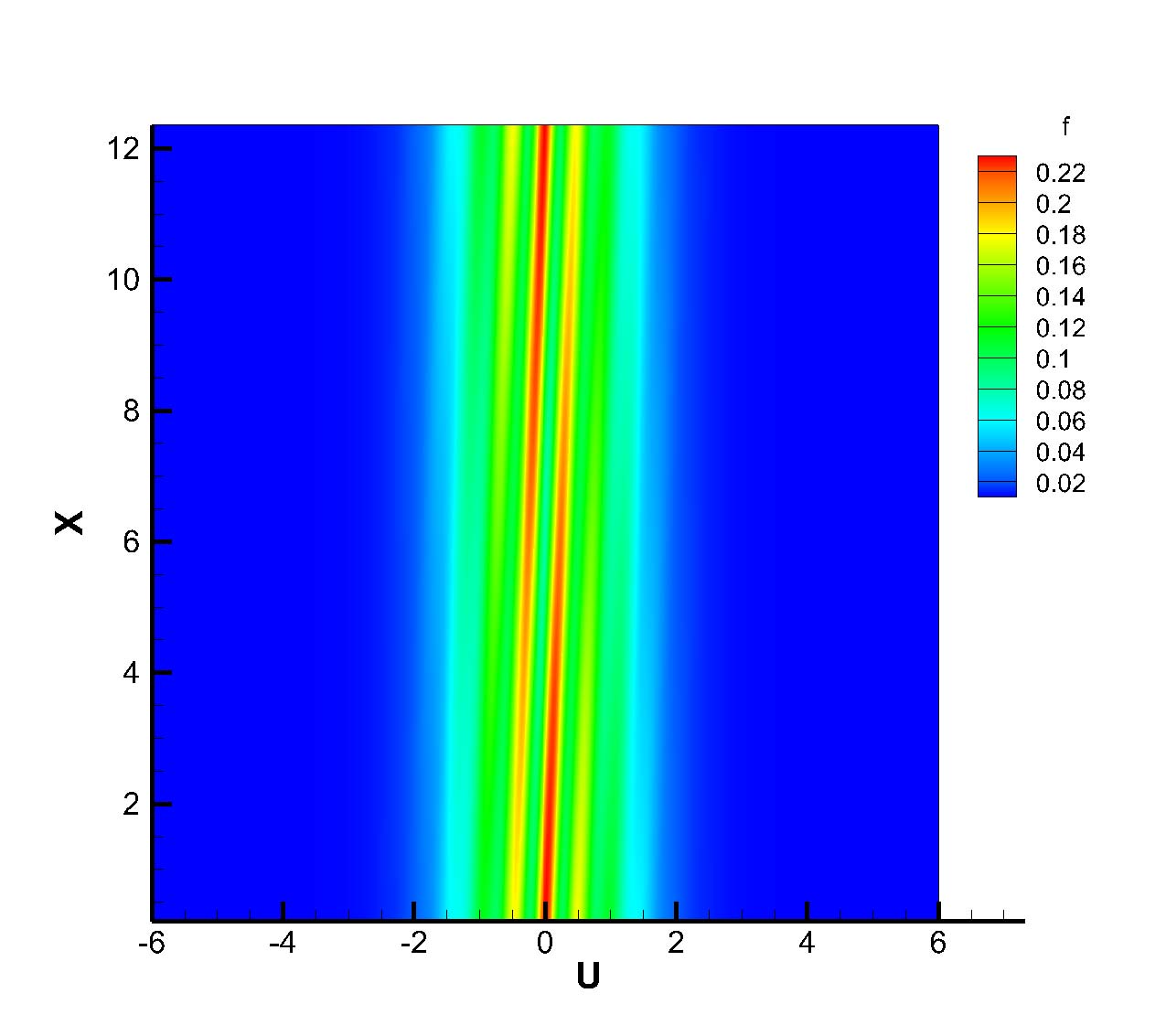}}
  \subfigure[$ t = 30$]{\label{fig:Fig09f}\includegraphics[width=0.45\textwidth]{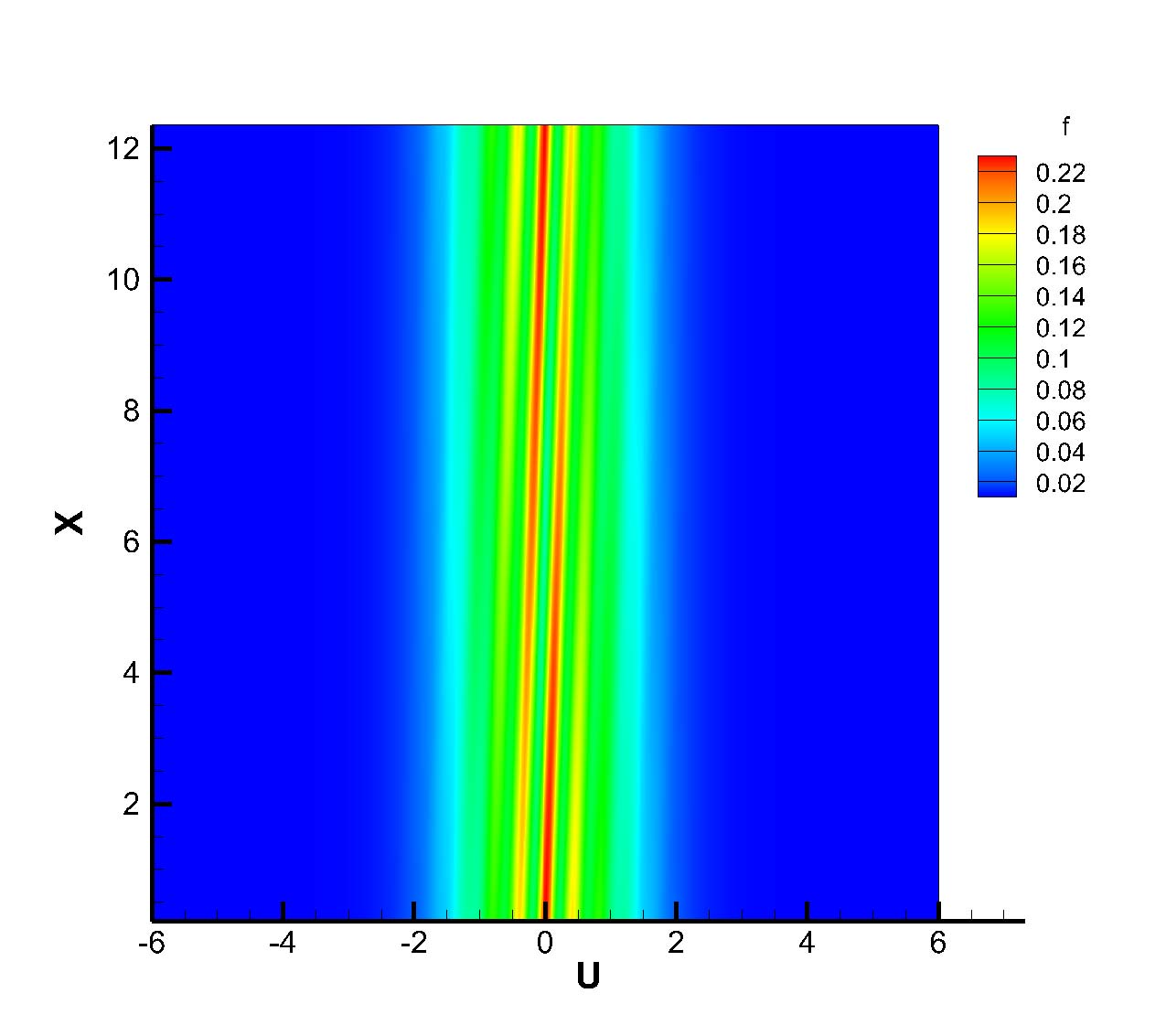}}
  \caption{Distribution functions in phase space at different time steps of nonlinear Landau damping.}
  \label{fig:Fig09}
\end{figure}

\begin{figure}
  \centering
  \subfigure[$ t = 5$]{\label{fig:Fig10a}\includegraphics[width=0.45\textwidth]{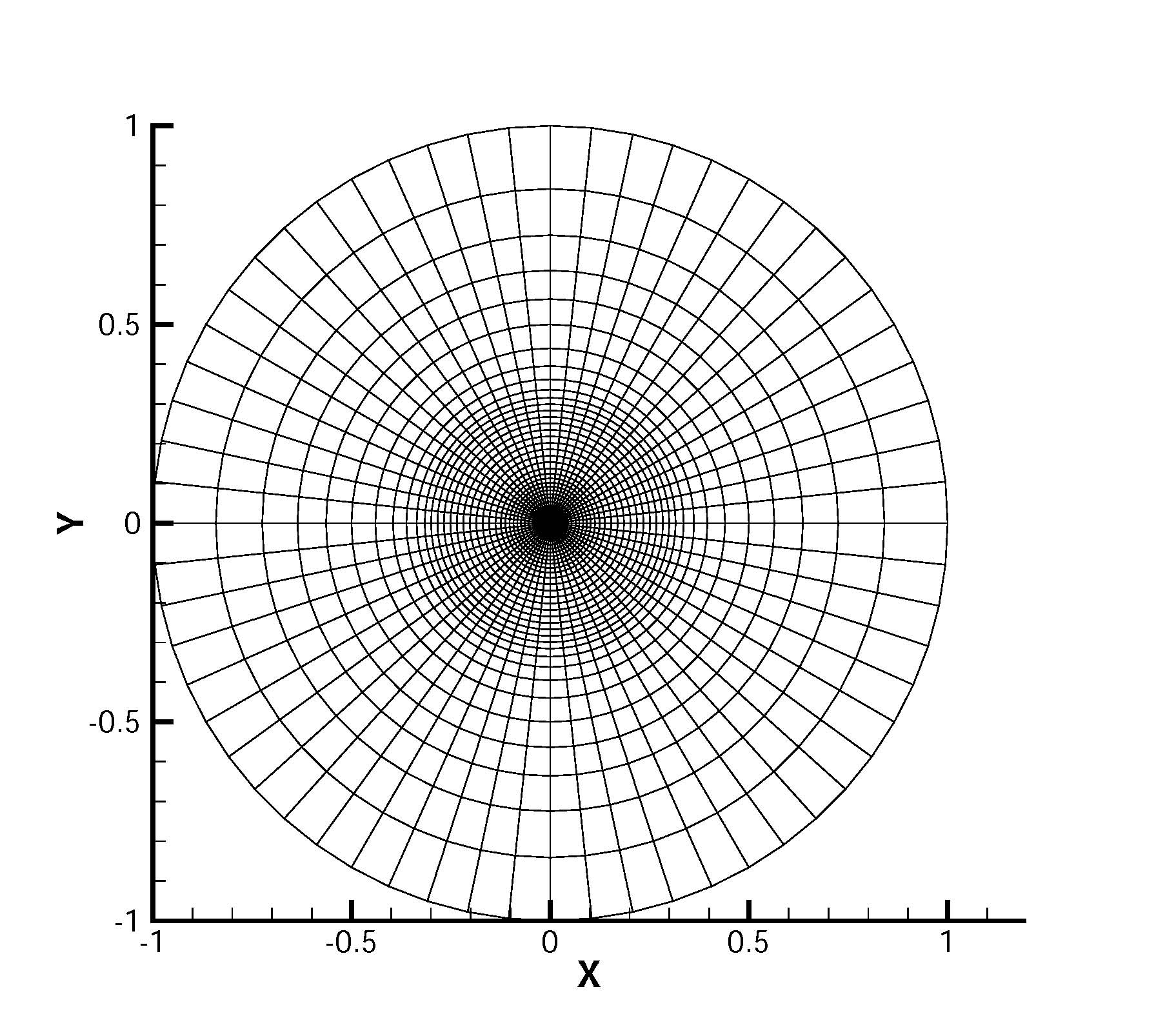}}
  \subfigure[$ t = 10$]{\label{fig:Fig10b}\includegraphics[width=0.45\textwidth]{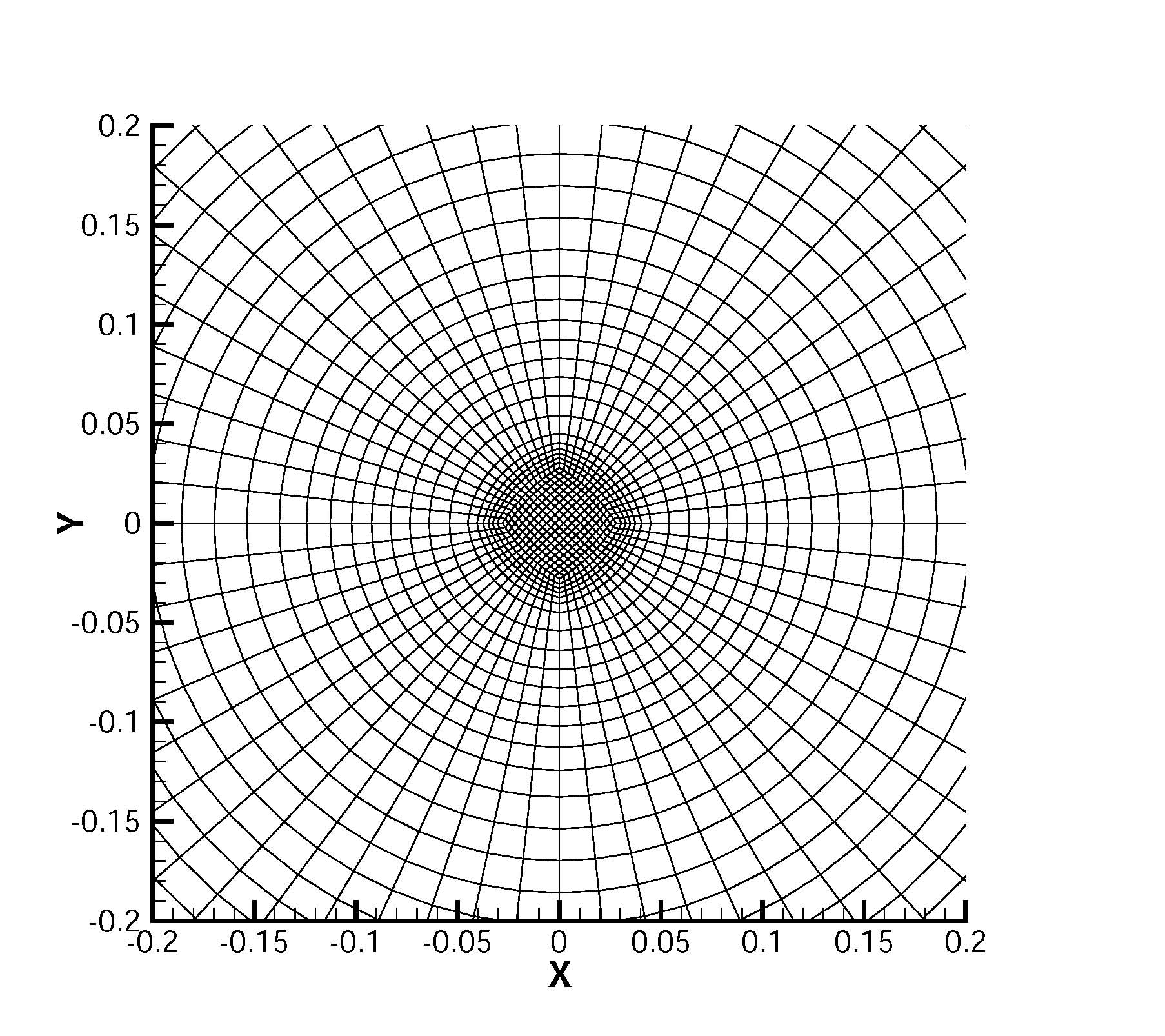}}
  \caption{Mesh for the simulation of the Gaussian beam.}
  \label{fig:Fig10}
\end{figure}

\begin{figure}
  \centering
  \includegraphics[width=0.5\textwidth]{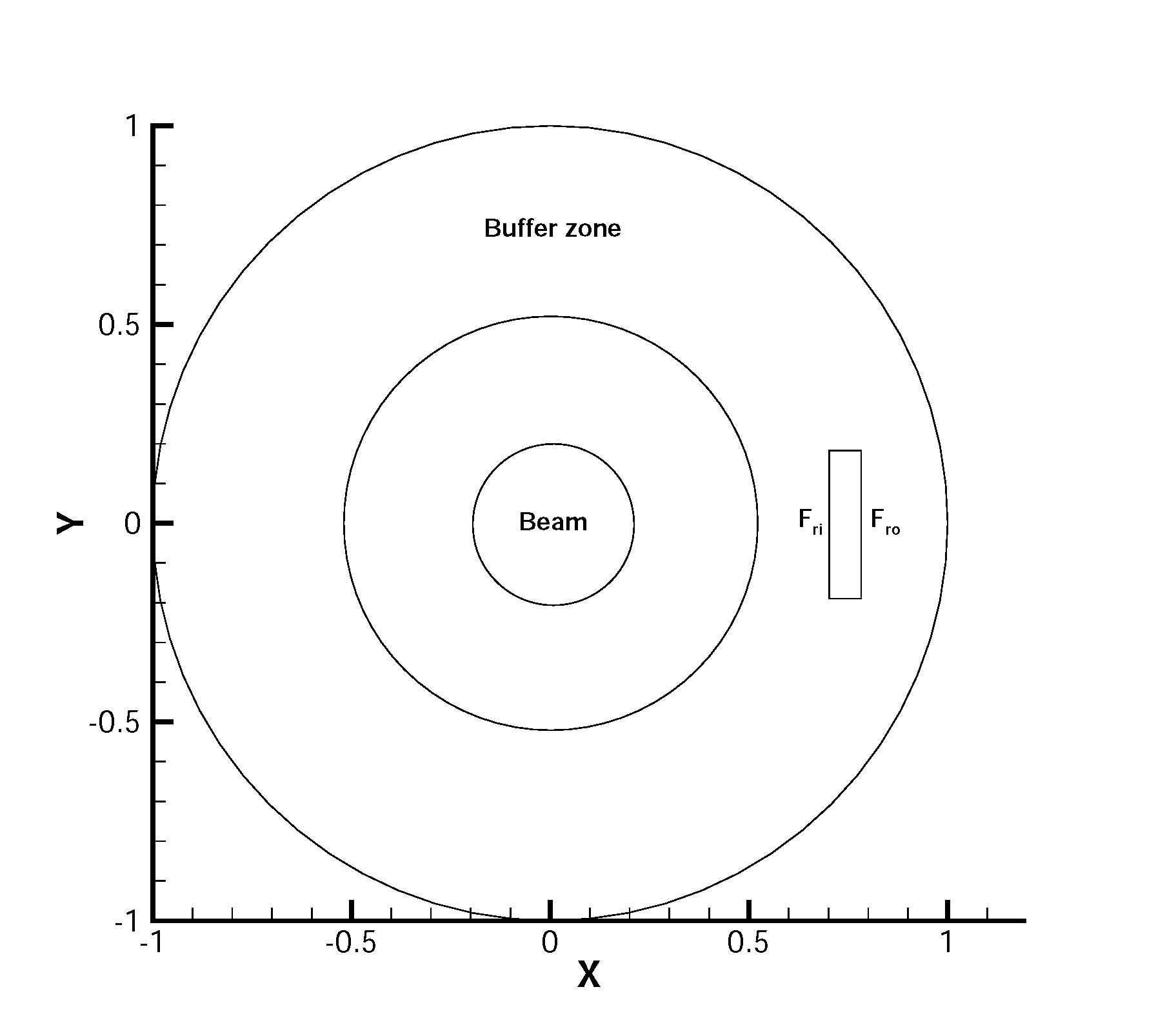}
  \caption{Buffer zone and flux on the warp for simulation of the Gaussian beam.}
  \label{fig:Fig11}
\end{figure}

\clearpage

\begin{figure}
  \centering
  \subfigure[$ t = 0.12$]{\label{fig:Fig12a}\includegraphics[width=0.45\textwidth]{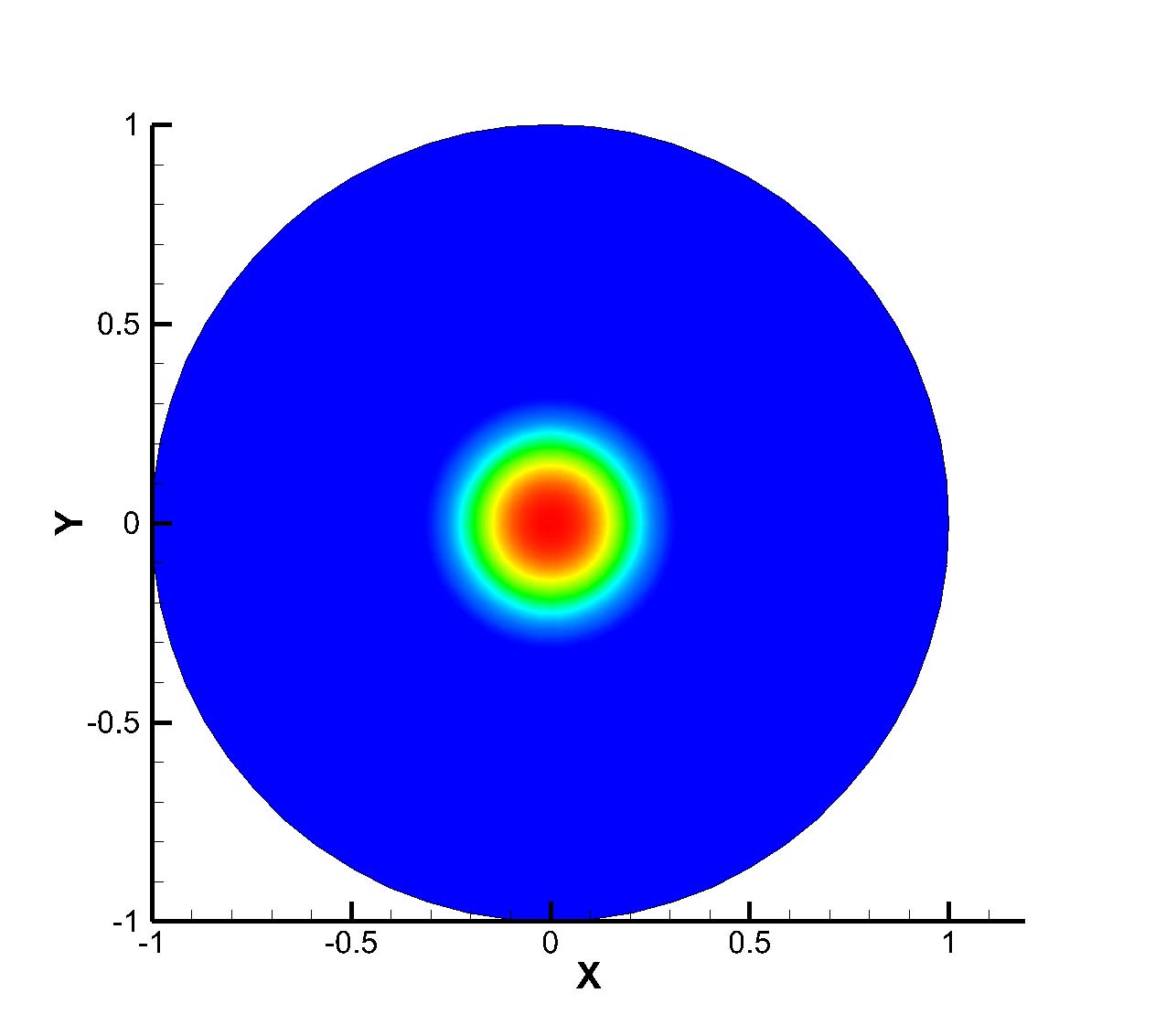}}
  \subfigure[$ t = 0.12$]{\label{fig:Fig12b}\includegraphics[width=0.45\textwidth]{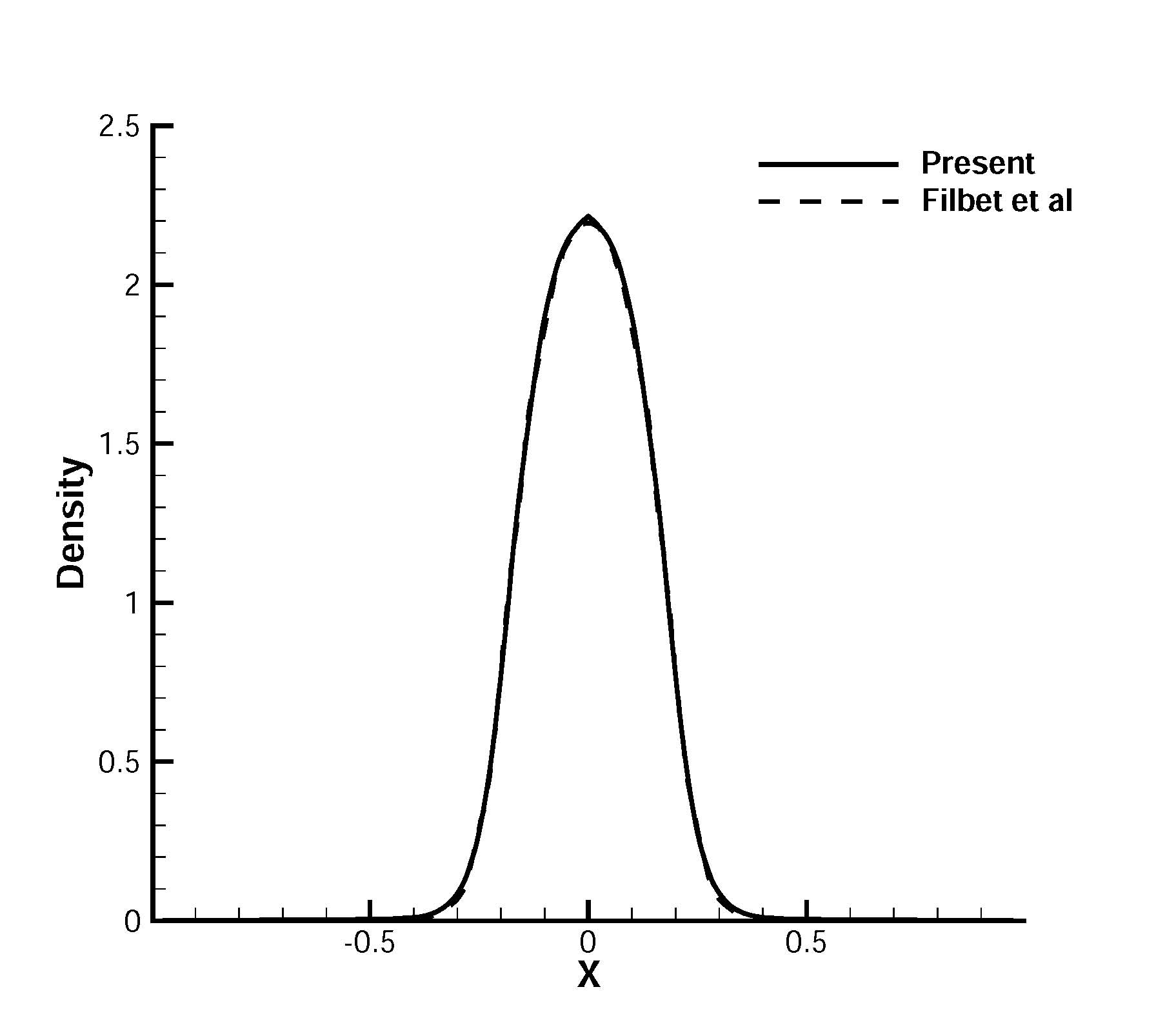}}
  \subfigure[$ t = 0.25$]{\label{fig:Fig12c}\includegraphics[width=0.45\textwidth]{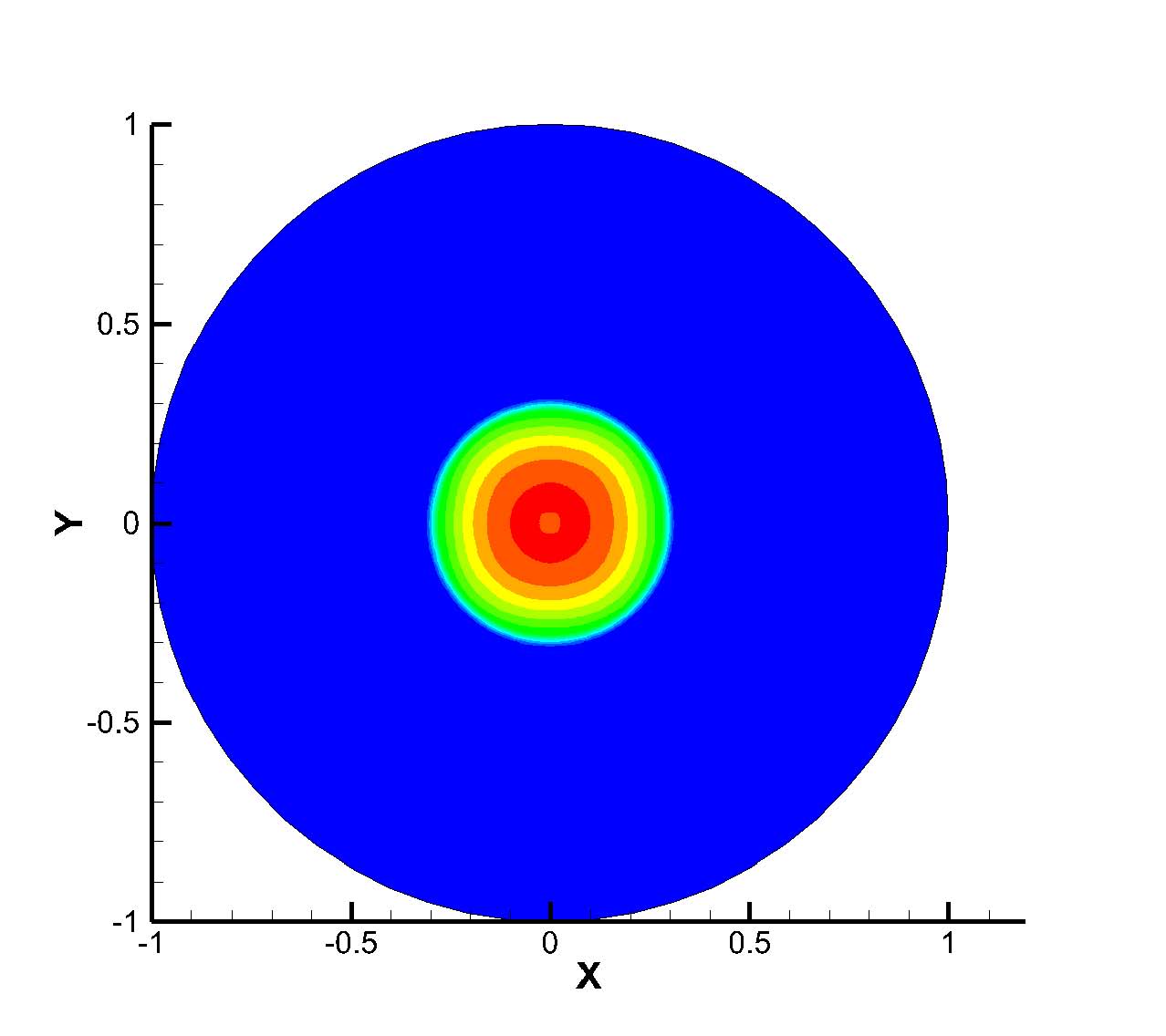}}
  \subfigure[$ t = 0.25$]{\label{fig:Fig12d}\includegraphics[width=0.45\textwidth]{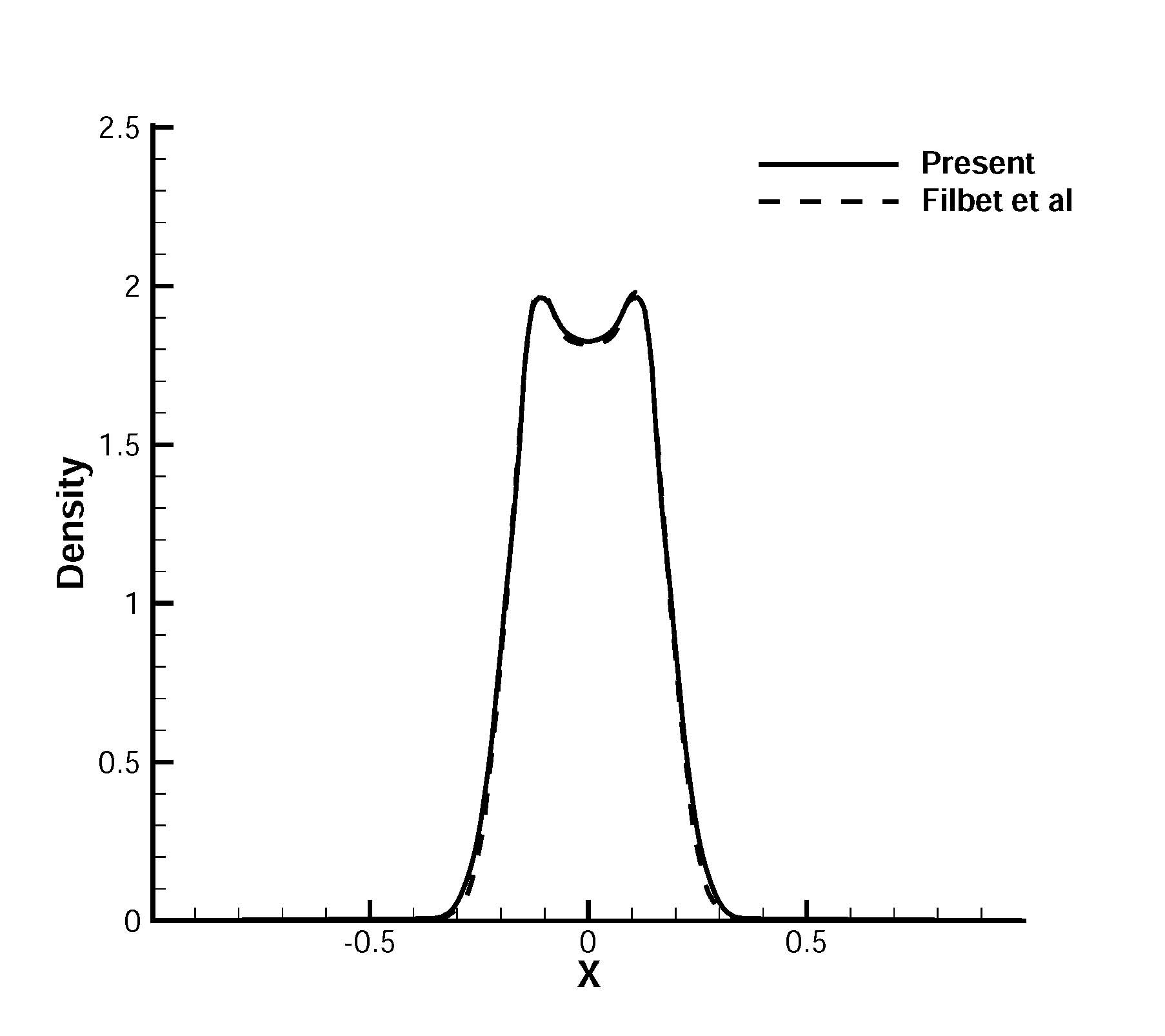}}
  \subfigure[$ t = 0.3$]{\label{fig:Fig12e}\includegraphics[width=0.45\textwidth]{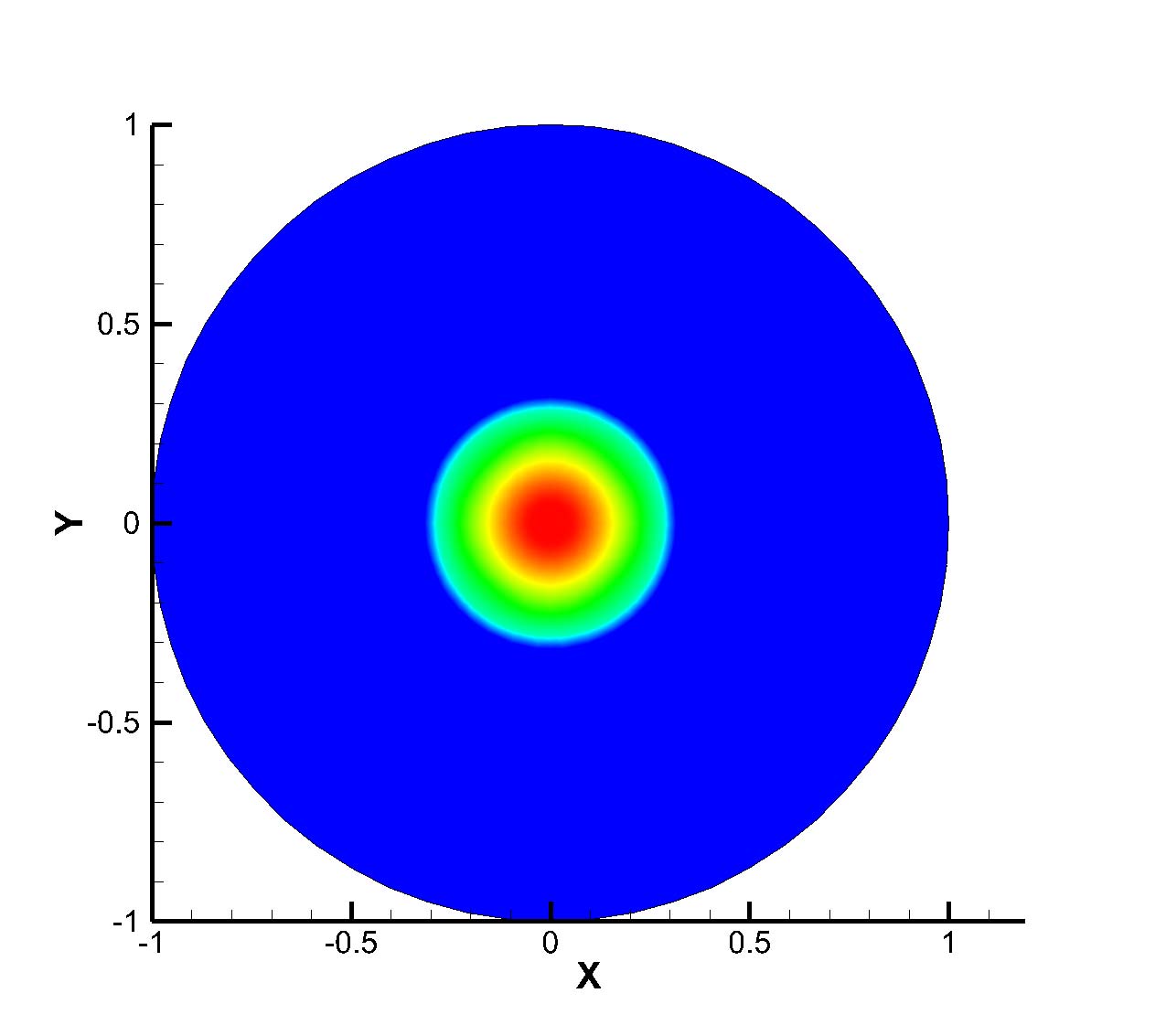}}
  \subfigure[$ t = 0.3$]{\label{fig:Fig12f}\includegraphics[width=0.45\textwidth]{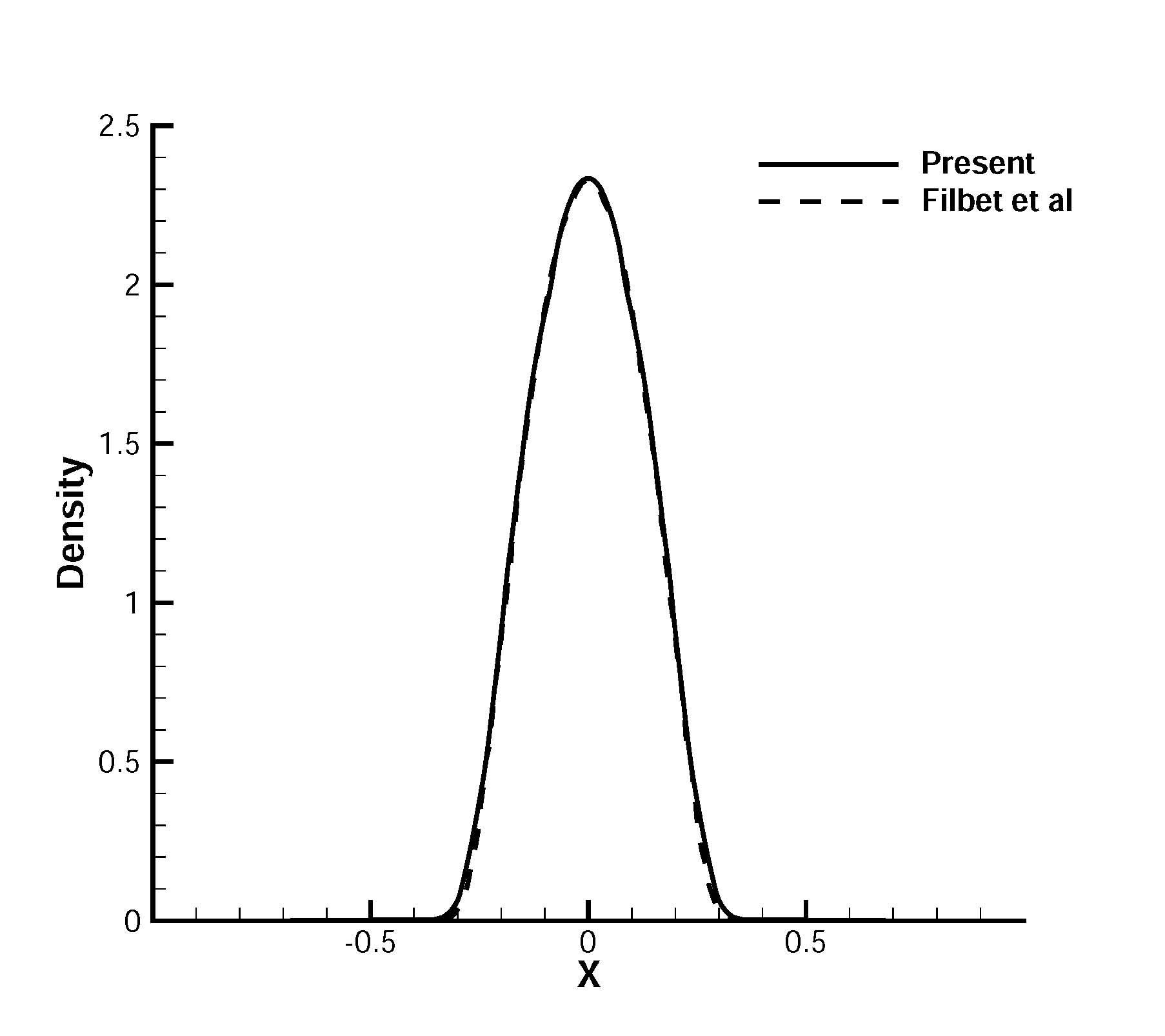}}
  \label{fig:Fig12}
\end{figure}

\begin{figure}
  \addtocounter{figure}{1}
  \centering
  \ContinuedFloat
  \subfigure[$ t = 0.5$]{\label{fig:Fig12g}\includegraphics[width=0.45\textwidth]{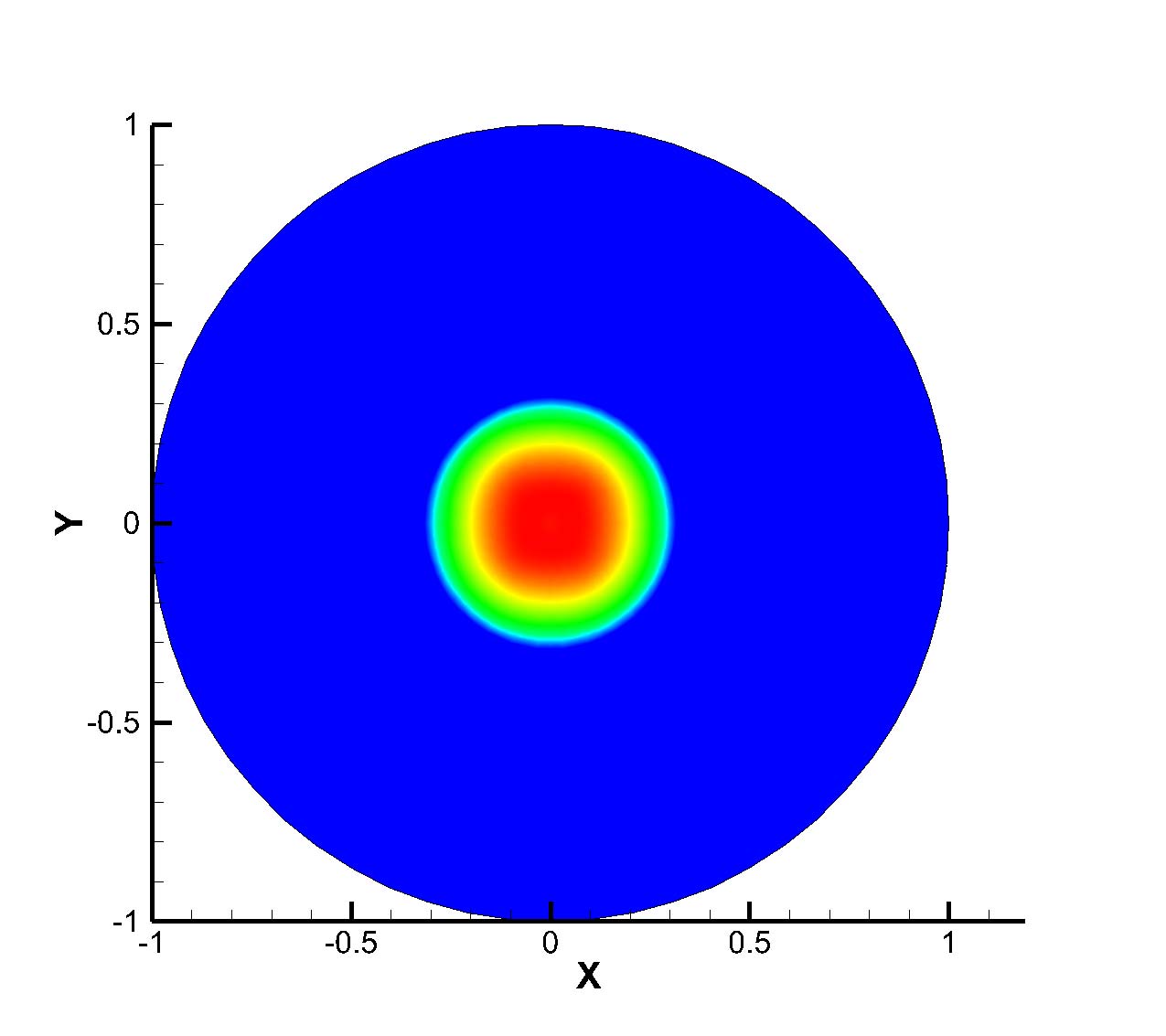}}
  \subfigure[$ t = 0.5$]{\label{fig:Fig12h}\includegraphics[width=0.45\textwidth]{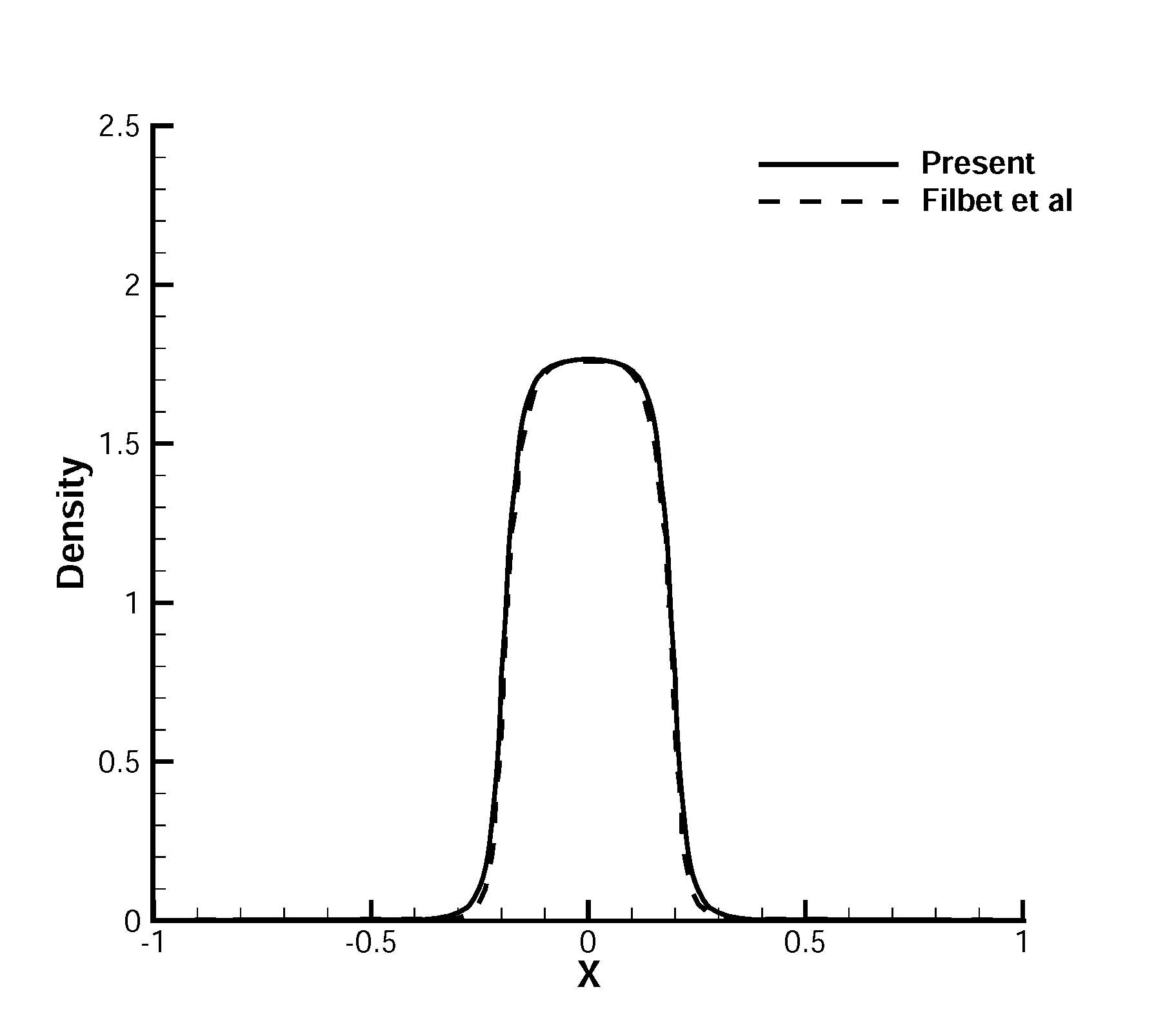}}
  \subfigure[$ t = 0.62$]{\label{fig:Fig12i}\includegraphics[width=0.45\textwidth]{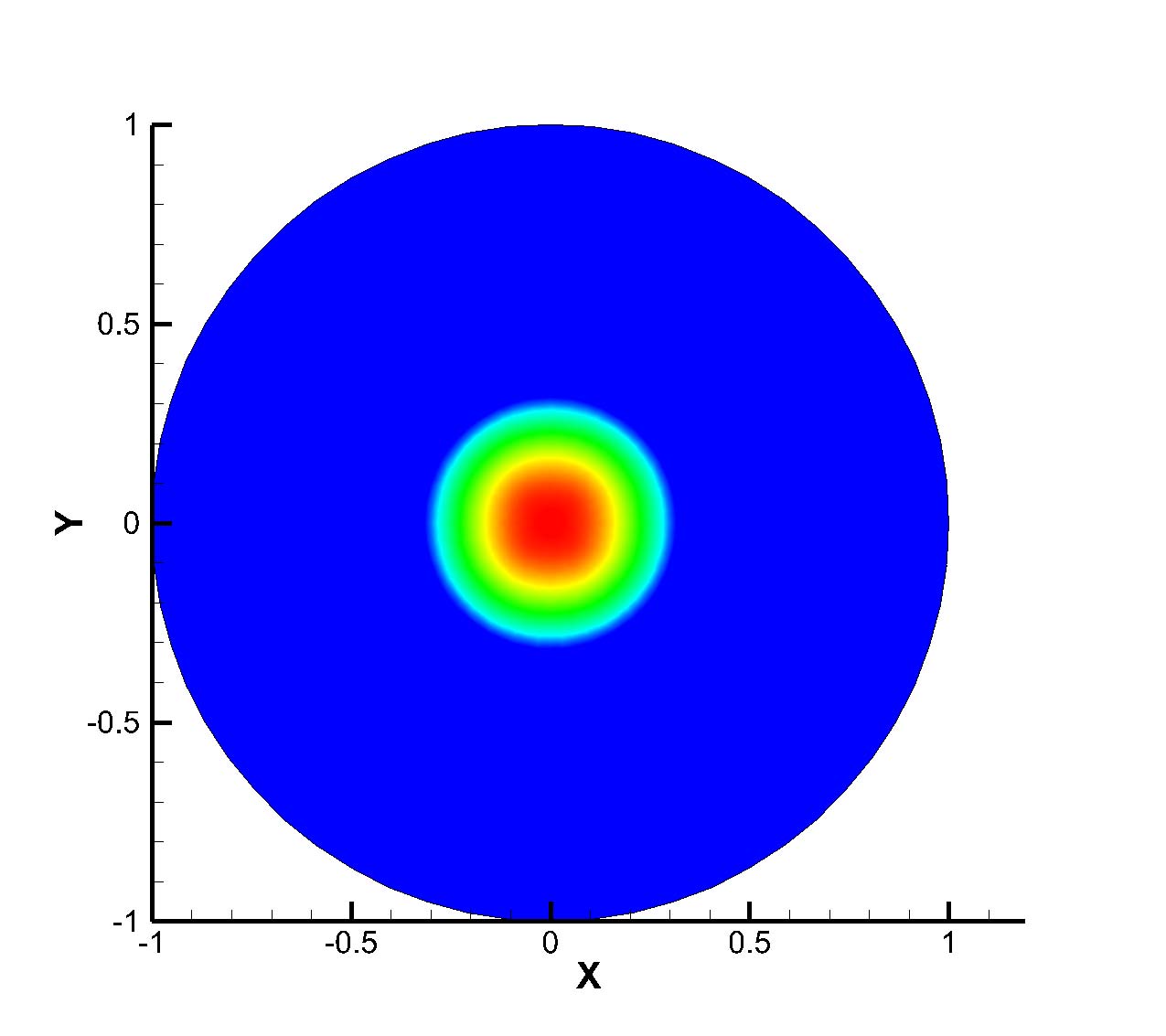}}
  \subfigure[$ t = 0.62$]{\label{fig:Fig12j}\includegraphics[width=0.45\textwidth]{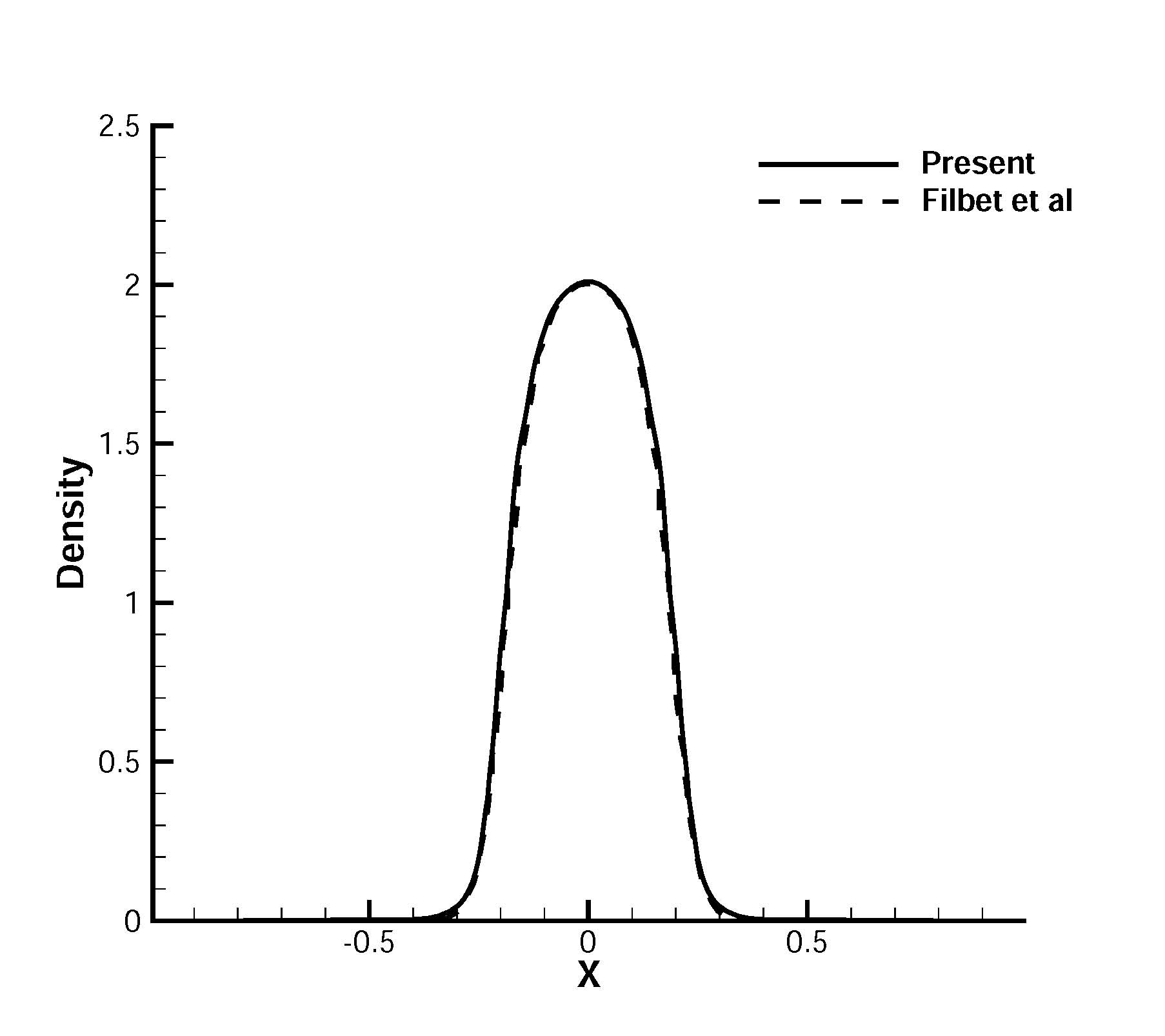}}
  \caption{Density distribution of the Gaussian beam at differenct time.}
  \label{fig:Fig12}
\end{figure}

\clearpage

\begin{figure}
  \centering
  \includegraphics[width=0.5\textwidth]{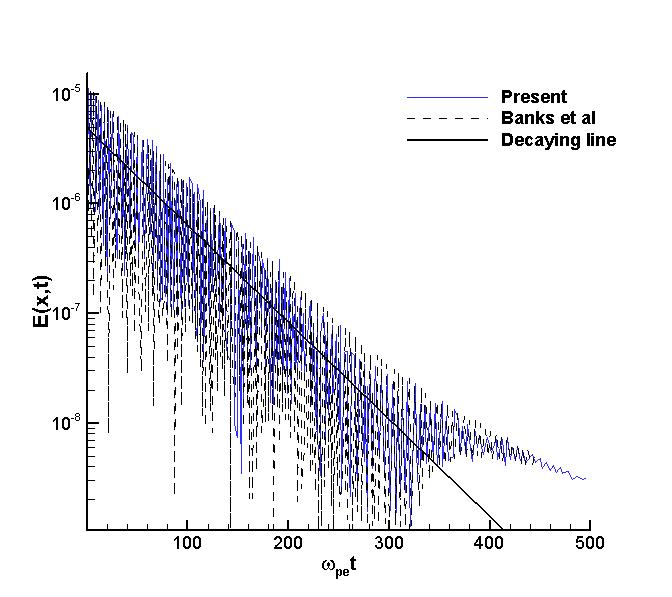}
  \caption{Amplitude of the perturbation's electric field.}
  \label{fig:Fig13}
\end{figure}

\end{document}